\documentclass[sigconf,screen]{acmart} %arxiv
\usepackage{natbib}

\newcommand{\N}{\textit{NexusAI}}

\newcommand{\CA}{\emph{Cognitive Abstraction}}
\newcommand{\WHV}{What-How-Value}

\usepackage{pifont}

\newcommand{\opr}[1]{\textcolor{gray}{\texttt{#1}}}

\usepackage{listings}

\usepackage{soul}% 高亮
\usepackage{subcaption}
\usepackage{xcolor}
\usepackage{algorithm}
\usepackage{algpseudocode}
\usepackage{caption}
\definecolor{set3a}{HTML}{00A896} % turquoise
\definecolor{set3b}{HTML}{D4A200} % pale yellow
\definecolor{set3c}{HTML}{7A6FF0} % lavender
\definecolor{set3d}{HTML}{F05A4F} % salmon
\definecolor{set3e}{HTML}{2F89D9} % light blue
\definecolor{set3f}{HTML}{F48C2A} % orange
\definecolor{set3g}{HTML}{63C74D} % light green
\definecolor{set3h}{HTML}{E85D9E} % pink
\definecolor{set3i}{HTML}{6B7285} % grey
\definecolor{set3j}{HTML}{A65FD4} % purple
\definecolor{set3k}{HTML}{3CBF8A} % mint
\definecolor{set3l}{HTML}{D4B800} % yellow

\AtBeginDocument{%
  }
\setcopyright{acmlicensed}
\copyrightyear{2018}
\acmYear{2018}
\acmDOI{XXXXXXX.XXXXXXX}

\acmConference[Conference acronym 'XX]{Make sure to enter the correct
  conference title from your rights confirmation email}{June 03--05,
  2018}{Woodstock, NY}

\acmISBN{978-1-4503-XXXX-X/2018/06}

\begin{document}

\title{NexusAI: Enabling Design Space Exploration of Ideas through Cognitive Abstraction and Functional Decomposition}
%NexusAI: Towards Design Space Exploration of Creative Ideas with Graph-based LLM
\renewcommand{\shortauthors}{Anqi Wang et al.}

\author{Anqi Wang}
% \email{awangan@connect.ust.hk}
\orcid{0000-0003-4238-6581}
\affiliation{%
  \institution{Hong Kong University of Science and Technology}
  \city{Hong Kong SAR}
  \state{}
  \country{China}
}

\author{Bingqian Wang}
% \email{bwang488@connect.hkust-gz.edu.cn}
\orcid{}
\affiliation{%
  \institution{Hong Kong University of Science and Technology (Guangzhou)}
  \city{Guangzhou}
  \state{Guangdong}
  \country{China}
}

\author{Huiyang Chen}
% \email{huiyangc@umich.edu}
\orcid{}
\affiliation{%
  \institution{University of Michigan}
  \city{Ann Arbor}
  \state{Michigan}
  \country{USA}
}

\author{Keqing Jiao}
% \email{}
\orcid{0009-0007-7115-4129}
\affiliation{%
  \institution{Carnegie Mellon University}
  \city{Pittsburgh}
  \state{Pennsylvania}
  \country{USA}
}

\author{Han Lei}
% \email{}
\orcid{}
\affiliation{%
  \institution{Hong Kong University of Science and Technology (Guangzhou)}
  \city{Guangzhou}
  \state{Guangdong}
  \country{China}
}

\author{Xin Tong}
% \email{xint@hkust-gz.edu.cn}
\orcid{0000-0002-8037-6301}
\affiliation{%
  \institution{Hong Kong University of Science and Technology (Guangzhou)}
  \city{Guangzhou}
  \country{China}}

\author{Pan Hui}
\authornote{Corresponding author}
\orcid{0000-0001-6026-1083}
\affiliation{%
  \institution{Hong Kong University of Science and Technology (Guangzhou)}
  \city{Guangzhou}
  \country{China}
  \email{panhui@hkust-gz.edu.cn}
}
% \affiliation{
%   \institution{Hong Kong University of Science and Technology}
%   \city{Hong Kong SAR}
%   \country{China}
%   \email{panhui@ust.hk}
% }

\begin{abstract}
    Large Language Models (LLMs) offer vast potential for creative ideation;  however, their standard interaction paradigm often produces unstructured textual outputs that lead users to prematurely converge on sub-optimal ideas—a phenomenon known as fixation.  While recent creativity tools have begun to structure these outputs, they remain compositionally opaque: ideas are organized as monolithic units that cannot be decomposed, abstracted, or recombinable at a sub-idea level.  To address this, we propose \textbf{Cognitive Abstraction (CA)}, a computational pipeline that transforms raw LLM-generated inspiration into a navigable and transformable design space.  We implement this pipeline in \textbf{NexusAI}, a prototype diagramming system that supports (I) \textbf{decomposition} of inspiration into typed functional fragments, (II) \textbf{multi-level abstraction} to externalize mental scaling, and (III) \textbf{cross-dimensional recombination} to spark novel design directions.  A within-subject user study ($N=14$) demonstrates that NexusAI significantly improves design space exploration, reduces cognitive overhead, and facilitates perspective reframing compared to a baseline.  Our work contributes: (1) a characterization of "compositional opacity" as a barrier in human-AI co-creation;  (2) the CA pipeline for operationalizing creative cognitive primitives at scale;  and (3) empirical evidence that structured, multi-level representations can effectively mitigate fixation and support divergent exploration.

\end{abstract}


% \ccsdesc[500]{Applied computing~Computer-aided design}

\begin{teaserfigure}
  \includegraphics[width=\textwidth]{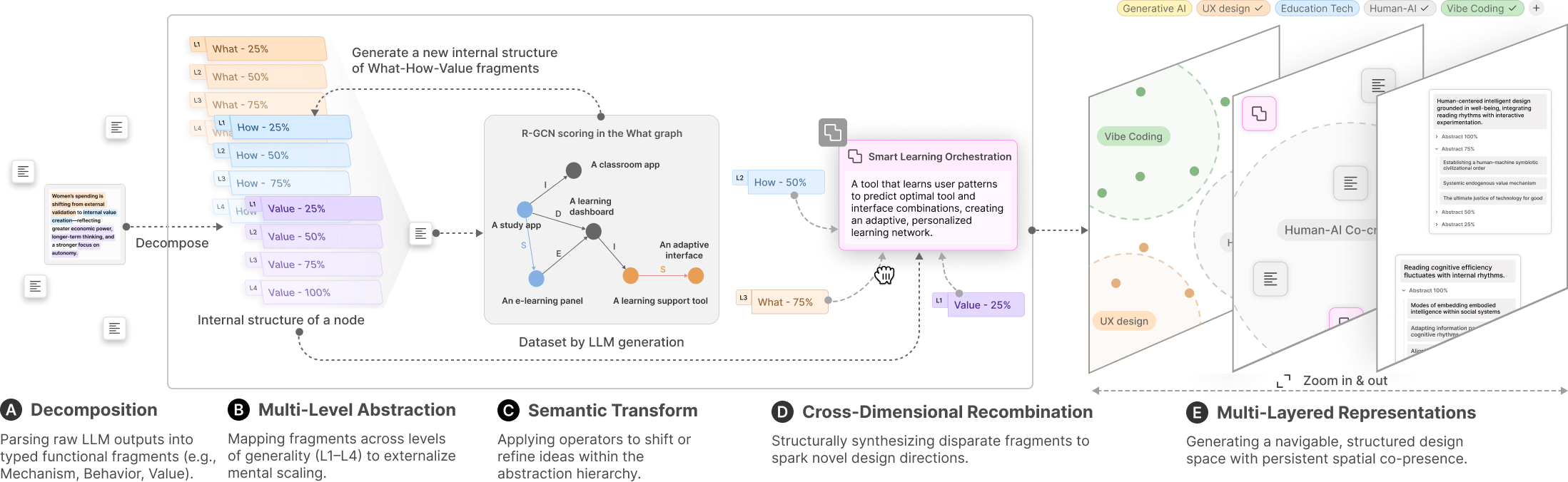}
  \caption{From Raw LLM Output to a Navigable Design Space via Cognitive Abstraction. Our approach addresses the gap in existing co-creation tools where ideas remain ``compositionally opaque.'' The pipeline (A-B) deconstructs and abstracts LLM-generated content into typed functional primitives, (C-D) provides a computational substrate for transforming and recombining these fragments at various abstraction levels, and finally (E) synthesizes them into a multi-view diagrammatic environment.   This workflow externalizes expert creative operations, facilitating fixation-resistant exploration through structural granularity.}
  \Description{}
  \label{fig:pipeline}
\end{teaserfigure}

\received{20 February 2007}
\received[revised]{12 March 2009}
\received[accepted]{5 June 2009}

\maketitle

\section{Introduction}

Recent advancements in Large Language Models (LLMs) have fundamentally reshaped the creative support tools, transitioning from simple prompt-response retrieval to facilitating the systematic exploration of complex \emph{design spaces} \cite{dove2016designspace,heape2007designspace,shaw2011designspaces,suh2024luminate,chen_coexploreds_2025}. By generating hierarchical idea structures and cross-domain concepts that often transcend a user's latent knowledge, LLMs offer a unique potential to catalyze ``sense-making'' in open-ended creative tasks\cite{chavula2022creative}. To steer this generative power, a new class of graph-based ideation interfaces has emerged, framing design space exploration as a navigable journey, moving beyond the limitations of ephemeral, unstructured text-based prompts. Systems such as \textit{IdeationWeb} \cite{shen_ideationweb_2025} and \textit{CoExploreDS} \cite{chen_coexploreds_2025} formalize the co-evolution of problems and solutions through structured nodes, while \textit{Luminate} \cite{suh2024luminate} enables exploratory navigation by projecting LLM outputs onto interactive dimensional maps. 
These tools collectively strive to reduce cognitive burden by providing a tangible structure for navigating multiple solution directions simultaneously.

However, a critical gap remains: despite their visual advantages, these systems often treat ideas as monolithic, indivisible units, resulting in what we define as \emph{compositional opacity}. Current interfaces typically present information as a ``black-box'' output, lacking an intermediate computational space that allows users to manipulate the internal mechanisms of an idea. Expert creative practice, conversely, demands fluid cognitive operations: (I) abstracting core mechanisms from their original contexts \cite{yee2019abstraction}, inferring underlying values from behavioral observations \cite{macneil2017dimensional}, and performing typed re-combinations of fragments across disparate dimensions. In existing workflows, these high-level operations must be performed entirely within the user's mind. At the peak of generative exploration, this places a severe strain on working memory \cite{baddeley2003working}, forcing a premature fixation on sub-optimal paths\cite{jansson1991design,cross2004expertise,dow2011parallel}. There is currently no computational pipeline that supports the cross-abstraction transformation of idea fragments or enables structured, cross-dimensional recombination—a missing link essential for true, interactive design space exploration. 

To bridge this gap, we propose Cognitive Abstraction (CA), an intermediate computational layer between monolithic LLM outputs and interactive design artifacts. Drawing on creative cognitive theories \cite{dorst2011core, yee2019abstraction} and our formative study (\S\ref{sec:formative}), the CA pipeline transforms raw inspiration into a navigable design space through three representational primitives: (1) Decomposition, parsing ideas into typed functional fragments (\textit{What-How-Value}) for independent manipulation \cite{alexander1977pattern}; (2) Abstraction, applying explicit operators to shift fragments across levels of generality, externalizing previously tacit mental shifts \cite{yee2019abstraction}; and (3) Cross-dimensional Operation, enabling the structured recombination of these functional parts into novel design directions \cite{hope_scaling_2022}. By exposing these primitives, we move the granularity of AI interaction beyond "whole ideas" to their constituent mechanisms, effectively offloading the cognitive strain of professional design synthesis. 
~
To demonstrate the feasibility and effectiveness, we initialize this pipeline in \N{} (Figure \ref{fig:pipeline}), an LLM-powered graphic diagramming tool featuring multi-level design space exploration. %The system operationalizes the CA primitives through a suite of interactive features, including semantic zooming for abstraction-level control and interest-based scaling to manage the visual density of the generated space. 

A within-subject study ($N$=14) shows that \N{}'s CA-driven workflow significantly reduces cognitive overhead and facilitates multi-directional exploration compared to baseline LLM interfaces. By providing persistent spatial co-presence of idea fragments, \N{} enables fluid perspective reframing and fixation-resistant synthesis. Our contributions include:
% Our primary contribution is not the invention of these individual strategies, but rather their systematic adaptation to the unique scale and heterogeneity of LLM-generated content. We contribute: 
\textbf{1. A Conceptual Framework:} The Cognitive Abstraction (CA) pipeline, formalizing a workflow that transforms monolithic LLM outputs into manipulable, multi-dimensional design spaces. %The Cognitive Abstraction (CA) pipeline, which identifies the core challenges of LLM-mediated ideation and formalizes a workflow for transforming raw model outputs into manipulable design spaces. 
\textbf{2. Interaction Mechanisms:} The design of \textit{What-How-Value} units and four-level abstraction operators (L1–L4), enabling typed, cross-dimensional operations on generative content at scale. %The design and implementation of specific mechanisms—including \textit{What-How-Value} (\WHV{}) manipulable units and four-level abstraction operators (L1–L4)—that enable users to perform typed, cross-dimensional operations at scale. 
\textbf{3. Empirical Validation:} A user study ($N$=14) demonstrating that externalizing these primitives mitigates design fixation and preserves the cognitive benefits of professional synthesis in AI workflows. %Evidence from a formal user study demonstrating how externalizing abstraction and decomposition effectively mitigates design fixation and preserves the cognitive benefits of professional creative practice within generative AI workflows.
In essence, we shift the granularity of Human-AI interaction from the ``whole idea'' to the ``constituent mechanism,'' fostering a more divergent and fixation-resistant exploration. 
% In essence, this work provides a blueprint for shifting the granularity of Human-AI interaction from the ``whole idea'' to the ``constituent mechanism,'' fostering a more divergent and fixation-resistant creative exploration. 

\section{Related Works}
    %\input{Sections/Related Works}
    %此版本为v1的减少字数版本
\label{sec:rw}
\subsection{Design Space Exploration and Cognitive Mechanisms} \label{sec:rw-cognitivemechanisms}

\emph{Design space} refers to the open-ended space of possibilities that emerges as users navigate, expand, and restructure problems and solutions~\cite{beaudouin2007prototyping,biskjaer2014constraint,dorst2011core,goel1995sketches,westerlund2005design,halskov2021filtering,macLean1991designsapce}. Rather than selecting from a predefined set, creativity arises from constructing and transforming the space itself~\cite{dove2017ux,heape2007designspace}---a process that is non-linear, iterative, and cognitively demanding. Central to this exploration are two cognitive operations: 
~
\emph{(I) Decomposition and Recombination:} Breaking complex ideas into manageable, ``typed'' primitives is foundational to creative synthesis~\cite{simon1991architecture, boden1991creativemind}. \citet{dorst2011core} formalizes this through the \textit{What-How-Value} logic, where creativity stems from mapping specific mechanisms (\emph{How}) to desired rationales (\emph{Value}). While recombining such primitives from distant domains is a key source of breakthroughs~\cite{chan2011benefits,fu2012meaning,gentner1997structure,gentner1998structure,cross2004expertise}, the process of ``chunking and parsing'' unstructured information imposes a significant cognitive load on users~\cite{miller1956magical, chunkingpharsing1995william}. 
\emph{(II) Abstraction:} This allows users to move across levels of specificity---from factual observations to structural principles~\cite{yee2019abstraction,liskov1986abstraction,mitprogrammingabstraction,suh_dynamic_2024}. High-level abstraction is essential for ``fixation-resistant'' exploration, enabling users to transfer mechanisms across disparate domains~\cite{moreno2015step,moreno2016overcoming, zahner2010fix}.

Despite their importance, these operations remain largely \emph{internal and tacit}. As LLMs increase the volume and heterogeneity of ideas, performing decomposition and abstraction mentally reaches a bottleneck, as working memory becomes overwhelmed by the ``compositional opacity'' of raw model outputs~\cite{baddeley2003working}. 

\subsection{Research on Ideation and LLM-Based Exploration}
%v1-paragraph1
Interactive ideation systems have long addressed creative exploration through \textbf{structured representations}---concept maps, design rationale graphs, dimension-based frameworks~\cite{macneil2017dimensional,subramonyam_solving_2022,lomas2021designspacecards}---and recombination mechanisms such as blending visual references~\cite{chilton2019visiblends,wang2023popblends}, icons~\cite{zhao2020iconate}, or stylistic elements~\cite{jeon2021fashionQ,chung2023artinter} and abstraction approaches \cite{yee2019abstraction,boggust_abstraction_2025,zahner2010fix}. In this way, visualization is an effective exploratory approaches that supports cross-source or cross-dimensional data \cite{card1999readings,watanabe2007bubble,hong2022scholastic,macneil2021probmap,palani2021conotate,zheng_disciplink_2024}, using multiple-view, such as cluster view \cite{cluster2015perteneder,wang2025cluster}. This strategy enables design space for visualize, navigate and manipulate cross-source or cross-dimensional data.

Recent LLM-based tools have adopted similar strategies to manage unstructured generative data (e.g.,~\cite{shen_ideationweb_2025, chen_coexploreds_2025, suh2024luminate,wang2026designerlyloop,jiang_graphologue_2023,themeviz2025kang,chung_patchview_2024,suh_sensecape_2023}). For instance, \textit{Luminate}~\cite{suh2024luminate} scaffolds design spaces via semantic dimensions, while \textit{Concept Induction}~\cite{lam_concept_2024} distills text into high-level concepts. \textit{Synthia}~\cite{zhang_synthia_2025} and \textit{Selenite}~\cite{liu2024selenite} further leverage LLMs for cross-domain sense-making through graphical interfaces. 
%v1-paragraph3
% \textcolor{blue}{In addition, current LLM-based works also cover wide domains that reflect the \textbf{cognitive mechanisms} for design space exploration. For instance, Synthia \cite{zhang_synthia_2025} and Selenite \cite{liu2024selenite}, leverages LLMs for cross-domain sense-making to navigate across sources or dimensions in query expansion, using interactive graphic-based approaches}~\cite{zhang_synthia_2025,jiang_graphologue_2023,themeviz2025kang,liu2024selenite,chung_patchview_2024}. Others utilize approaches including high-level abstraction~\cite{lam2024concept}, chunking and decomposition~\cite{wang2023popblends,choiCreativeConnectSupportingReference2024}. \aq{for instance} 
However, %as summarized in Table \ref{tab:representationmethods},
a critical gap remains: these systems predominantly operate on \emph{whole references} or end results. They provide limited support for isolating and re-configuring the intermediate ``cognitive artifacts'' %(the \WHV{} fragments) 
that professional designers actually reason with. None provides a unified computational pipeline that externalizes the transition across abstraction levels or supports typed, cross-dimensional recombination of fragments. 
This gap motivates our \emph{Cognitive Abstraction} pipeline, which treats these cognitive operations as interactive primitives.
%v1-paragraph4
% This gap motivates the \emph{Cognitive Abstraction} pipeline we develop: an intermediate computational layer between raw LLM output and interactive design artifacts, grounded in Decomposition, Abstraction, and Cross-dimensional operation as first-class interactive primitives.

\section{Formative Study}
    \label{sec:formative}
% \hl{To understand how users explore design spaces and to inform system design, we conducted a two-part formative study with participants who have LLM-supported creativity experience. 
% We focused on identifying current practices, challenges, and expectations for LLM-assisted design tools.} The studies were approved by the university IRB. 
To understand how users explore design spaces and to inform system design, we conducted two IRB-approved formative studies with LLM-experienced creators to identify current practices, challenges, and expectations for LLM-assisted design tools. 

\subsection{Study 1: Understanding Exploring Design Space Using LLMs}
%before20Mar
\textit{Participants.} To understand how designers explore design space with LLMs, we conducted semi-structured interviews with eight participants (5 female, 3 male; aged 25–30) recruited via purposeful sampling from professional design networks and university departments. Inclusion criteria required at least one design project experience and one year of weekly LLM use; participants provided full records of prior LLM interactions. Informed consent was obtained online; participation was voluntary and uncompensated (see Appendix, Table~\ref{tab:fs1-participants}).

\textit{Procedure and Analysis.} Interviews covered (1) backgrounds and usage of existing LLM-assisted creativity tools, (2) prior experience with AI-assisted design, (3) expected interaction needs, and (4) visions for future tools. Data were analyzed via open coding~\cite{corbin2014basics}: two researchers independently coded transcripts, resolved discrepancies through calibration meetings to consolidate a shared codebook, and developed higher-level categories by aggregating related codes---with a senior researcher reviewing for methodological rigor. 

\subsubsection{Findings} 
Our formative study with seven designers revealed three interconnected challenges in LLM-assisted creative exploration. First, designers exhibited systematic \textbf{cognitive fixation} through premature convergence on initial outputs, exacerbated by LLMs' inability to externalize the implicit design space or differentiate non-redundant insights within monolithic text responses (\textbf{C1}). 
Second, the absence of \textbf{decomposable semantic units} prevented designers from performing the comparative and recombinatorial operations essential to creative exploration, leaving potentially generative cross-domain correspondences tacit and unexploited (\textbf{C2}). 
Third, the lack of \textbf{vertical abstraction scaffolding}---spanning decontextualization, mechanistic inference, and analogical value transfer---confined exploration to lateral variation at a single level of specificity, precluding the conceptual leaps that drive creative breakthroughs (\textbf{C3}). Together, these challenges reveal a structural gap between LLMs' generative capacity and designers' need for navigable, manipulable, and multi-level design spaces.

% \textcolor{gray}{\textbf{The central thesis of this paper is that:} These challenges collectively arise from treating LLM-generated ideas as ephemeral, atomic outputs rather than as \textit{structured, persistent design elements} that can be decomposed into semantic components (What-How-Value), organized across abstraction levels, and recombined through explicit graph operations. Addressing this requires a shift from conversation-based interaction to \textit{graph-based design space construction}, where AI supports not just generation but the systematic externalization and manipulation of conceptual structures underlying creative exploration.}

\subsection{Study 2: Investigating Interaction Approaches} 
% To identify approaches addressing C1--C3, we conducted a literature survey and design workshops, grounding our interaction approaches in three operations: \emph{Decomposition}~\cite{dorst2011core} (C1, C2), \emph{Abstraction}~\cite{yee2019abstraction,liskov1986abstraction,zahner2010fix,alexander1977pattern} (C1, C3), and \emph{Cross-Dimensional Operation} (C2, C3) (details in Section \ref{sec:rw}).
To identify strategies for addressing C1--C3, we conducted a literature survey and a series of design workshops. This process grounded our interaction framework in three core operations: (1) Decomposition~\cite{dorst2011core} (C1, C2); (2) Abstraction~\cite{yee2019abstraction, liskov1986abstraction, zahner2010fix, alexander1977pattern} (C1, C3); and (3) Cross-Dimensional Operation (C2, C3). Detailed theoretical underpinnings and related work are provided in Section \ref{sec:rw}.

% ~~~~~~~~~~~~~~~~~~~~~~
\textit{Workshop Protocol \& Participants}
To validate these approaches in real usages, we conducted design workshops with six creative practitioners (4 female, 2 male; aged 25–30) recruited via purposeful sampling. Participants (P1--P6) represented diverse design domains (UIUX, product, visual, industrial, HCI, architectural) with 1--8 years of creative experience and regular LLM usage (Appendix, Table~\ref{tab:fs2-participants}).

\textit{Task and Procedure} 
After informed consent, participants accessed a FigJam board containing 20 domain-agnostic information snippets. The study comprised three phases: (1) \emph{Browsing (10 min)}: examining all items; (2) \emph{Design Exploration (30 min)}: manipulating information via sticky notes on a FigJam (see Appendix, Figure \ref{fig:study2}), requiring participants to browse notes and brainstorm by decomposed \WHV{} fragments, and coming up with new ideas; (3) \emph{Interview (20 min)}: describing their process via semi-structured interview. 
The semi-structured interview probed participants’ (1) design process and idea evolution, (2) interaction with \WHV{} fragments, (3) use of abstraction and decomposition, and (4) comparisons with their conventional workflows. We further elicited rationales behind key decisions and gathered reflections on usability and potential improvements. 
Think-aloud protocol~\cite{Ericsson1980VerbalRA} was encouraged and all sessions were screen and audio recorded.

\textit{Data Analysis} 
Transcribed sessions underwent two-phase analysis. \emph{Phase 1:} Theory-informed deductive coding~\cite{hsieh2005three} decomposed annotations into \WHV{} components (definition in \S\ref{sec:rw-cognitivemechanisms})—and mapped each to one of four abstraction levels based on a predefined codebook: 
\emph{L1-Fact: facts, data, observable behavior;}
\emph{L2-Insight: patterns, drivers, structural relations;}
\emph{L3-Principle:  design principles, transferable values;}
\emph{L4-Vision: future narratives, systemic framings.}
~
\emph{Phase 2:} Sequential interaction analysis~\cite{jordan1995interaction} identified recurrent configurations across temporal contexts. Two researchers independently coded with consensus reached through evidence-based deliberation.

~~~~~~~~~~~~~~~~~~~~~~~~~~~~~

% \subsubsection{Findings.}\label{sec:fs-study2-findings}
% \textbf{I1. Structuring Representations Spatially and Chronologically (C1).}
% Users used graphical proximity to organize cross-thematic information iteratively and nonlinearly, using spatial arrangement as an external scaffold to offload working memory. 

% \textbf{I2. Decomposing and Abstracting Information (C2, C3).}
% Users deconstructed raw information into typed components, transforming complex semantics into clear, independently manipulable units, based on \WHV{} units. Three abstraction approaches emerged: \emph{Decontextualized Elevation (What);} \emph{Mechanistic Inference (How);} and \emph{Analogical Transfer (Value)} (details in Appendix \ref{apx:fs-three_abstraction}), covering operations on \WHV{} fragments at different abstraction levels (L1--L4), confirming that abstraction is not a single operation but a typed, multi-path process. 

% \textbf{I3. Cross-Dimensional Fragment Manipulation (C2, C3).}
% Participants recombined abstract fragments across \WHV{} as manipulable units to generate design concepts in nonlinear patterns. 
% Users recombined abstract fragments across \WHV{} fragments in nonlinear patterns by recombining any two typed fragment, rather than the linear ``Value+How$\rightarrow$What'' proposed by~\citet{dorst2011core}. In addition to this flexible recombination, 
% four recurrent cross-dimensional modes emerged, we illustrated details in \S\ref{sec:pip_S2}. 

\subsubsection{Design Goals} 
Study 1 and Study 2 together surface a shared structural gap: users naturally perform \emph{Decomposition}, \emph{Abstraction}, and \emph{Cross-dimensional operation} as core cognitive processes, yet LLM interfaces provide no computational substrate to externalize or scaffold these operations. This motivates the three design goals: 

\textbf{DG1. Provide Structured, Multi-Layered Representations to Externalize and Organize Fragmented Information (C1, C2, C3).}
The system should provide explicit representational grammars—\WHV{} triplets, thematic clusters, relation links, and abstraction-level labels (L1--L4)—to help users externalize and organize fragmented inspiration into persistent, navigable structures. This directly addresses the monolithic opacity of LLM outputs (C2), the absence of graph-based representations that make cross-domain correspondences explicit (C3), and users' reliance on working memory to track novelty and relevance across turns (C1). 

\textbf{DG2. Enable Multi-Level Abstraction and Perspective Transformation to Broaden the Design Space (C3).}
The system should provide abstraction scaffolds that transform \WHV{} fragments vertically—from generalization (L3/L4) to concretization (L1/L2)—and across alternative perspectives, revealing new opportunities and reducing fixation. As participants' usage patterns confirmed, abstraction is not a single operation but a typed, multi-path process (\emph{Decontextualized Elevation}, \emph{Mechanistic Inference}, \emph{Analogical Transfer}), each operating on a distinct \WHV{} role. 

\textbf{DG3. Support Structured Cross-Dimensional Recombination to Generate Divergent Design Directions (C2, C3).}
The system should treat cross-dimensional recombination as a core generative capability, offering structured combination of \WHV{} fragments across functional roles in nonlinear patterns to produce multiple alternative design directions and counteract premature convergence. 

\section{System Design}
    \label{sec:systemdesign}

\begin{figure*}
    \centering
    \includegraphics[width=0.8\linewidth]{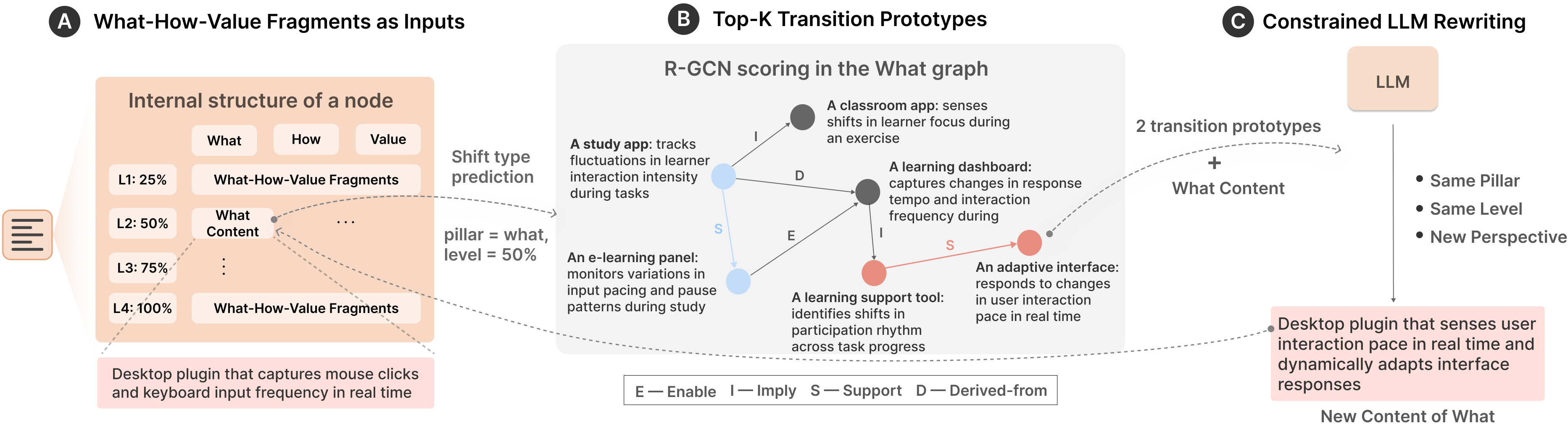}

    \caption{The R-GCN-guided \WHV{} rewriting mechanism. The process initiates with a structured \WHV{} fragment extracted from a specific node (A). Subsequently, an R-GCN scoring module predicts the transition type and retrieves the Top-$K$ prototypes from the relevant pillar-specific graph (B). These prototypes are then integrated into the LLM prompt as structural constraints (C). This pipeline facilitates the generation of a refined fragment that maintains the original pillar and abstraction level while introducing a novel perspective. Although the figure illustrates this workflow using a What fragment at the L2 (50\%) level, the mechanism is uniformly applicable to all What, How, and Value fragments across all levels of abstraction.}
    \label{fig:rgcn_whv_rewriting}
\end{figure*}
We present \N{}, a graph-based system that operationalizes cognitive design processes. Its core \emph{Cognitive Abstraction (CA) pipeline} transforms unstructured inspiration into a structured, navigable design space across stages corresponding to DG1--DG3 (Figure~\ref{fig:pipeline}). 
% We present \N{}, a graph-based LLM interaction system that operationalizes the cognitive processes% identified in our formative studies into a computational framework
% . At its core is the \emph{Cognitive Abstraction (CA) pipeline}—an LLM-driven reasoning system that transforms unstructured inspiration into a structured, navigable design space through several stages corresponding to DG1--DG3 (Figure~\ref{fig:pipeline}). 

\subsection{\CA{} Pipeline}\label{sec:pip} 
This \CA{} pipeline is across five stages. 
\subsubsection{Stage 1: \WHV{} Extractor (DG1)} \label{sec:pip_S1}
Raw inspiration—user notes, LLM outputs, case studies—arrives as monolithic text that resists decomposition and cross-dimensional manipulation (C1, C2) (Figure \ref{fig:pipeline}A--B). To make this material computationally tractable, Stage~1 parses each input into atomic, type-labeled fragments structured by the \WHV{} schema (drawing on literature \cite{dorst2011core}), producing the typed units that all downstream stages operate on.
 
The extractor runs a two-step LLM inference: first decomposing the input into elementary meaning units, then assigning each unit a \WHV{} fragment (\emph{What}: artifact or problem; \emph{How}: mechanism or strategy; \emph{Value}: rationale or criterion) and an abstraction level (L1--L4). Role definitions and a few-shot pool of valid \WHV{} patterns are embedded in the prompt to enforce schema conformance; post-processing further constrains outputs to the predefined role-and-level boundaries (full prompt and definitions in Appendix~\ref{apx:whv-definition}, \ref{apx:whv-fewshot}, \ref{apx:prompt-WHVextractor}). The resulting fragments serve as the manipulable units for \opr{Analyze}, \opr{Transform}, and \opr{Merge} interactions throughout the system.

\subsubsection{Stage 2: \WHV{} Semantic Transform (DG2)} \label{sec:pip_S2}
Study~2 identified three role-specific abstraction mechanisms (Figure \ref{fig:pipeline} C). 
These mechanisms share a common constraint: the perspective shift must preserve the fragment's original \WHV{} role and abstraction level (L1--L4), producing a structurally equivalent but semantically reframed expression. 

A naive approach would prompt an LLM directly to perform such rewriting. However, unconstrained LLM generation tends to produce arbitrary paraphrases or level-drifting outputs that violate the \WHV{} schema. The key insight is that perspective shifts of this kind are \emph{not arbitrary}: they follow learnable structural transition patterns observable in real design reasoning examples. To ground and constrain the rewriting process, we design an \emph{R-GCN-guided \WHV{} rewriting mechanism} that retrieves structurally matched transition prototypes from an offline graph and injects them as constraints into LLM generation (Figure~\ref{fig:rgcn_whv_rewriting}).

\paragraph{Offline Prototype Graph Construction.}
%润色后的
We pre-construct a multi-relational prototype graph $\mathcal{G} = (\mathcal{V}, \mathcal{E}, \mathcal{R})$, where $\mathcal{V}$ denotes the set of \WHV{} fragment expressions and $\mathcal{R}$ represents the set of transformation types. These types are derived from the FBS (Function--Behaviour--Structure) framework \cite{gero1990design, gero2004situated} and reflect the transformation tendencies observed in user seed examples. The set of directed labeled edges, $\mathcal{E} \subseteq \mathcal{V} \times \mathcal{R} \times \mathcal{V}$, encodes observed $(\textit{source fragment, transformation type, target fragment})$ transitions. To maintain structural consistency, the graph is partitioned into three pillar-specific subgraphs (\WHV{}), with transitions further stratified by abstraction level. 
Unlike graphs induced from general web corpora, this prototype graph is constructed offline from a curated set of 20 transformation examples certified by two experts, derived from prior literature \cite{dorst2011core}. This core dataset was subsequently expanded using GPT under explicit \WHV{} constraints. We initially generated approximately 1,000 transformation samples per pillar; following deduplication and validity filtering, over 900 effective samples were retained for the final graph construction.
%bq版本
% We pre-build a multi-relational prototype graph $G=(V,E,R)$, where $V$ is the set of \WHV{} fragment expressions, $R$ is \textcolor{black}{the set of shift types adapted for \WHV{} fragments, derived from the transformation tendencies reflected in user seed examples and the canonical transformation logic revealed in the FBS framework\cite{gero1990design,gero2004situated} (Function--Behaviour--Structure)}, and $E \subseteq V \times R \times V$ is the set of directed labeled edges, each encoding an observed \textit{(source fragment, shift type, target fragment)} transition. The graph is partitioned into three pillar-specific subgraphs corresponding to \emph{What}, \emph{How}, and \emph{Value}, with transitions further stratified by abstraction level to ensure \textcolor{black}{structural consistency}. 
% \textcolor{black}{Rather than being directly induced from general web corpora, the graph is constructed offline from a small set of user-derived transformation examples, informed by transformation exemplars reported in prior work, and then expanded with GPT under explicit \WHV{} constraints. For each pillar, we initially constructed around 1{,}000 transformation samples; after deduplication and validity filtering, more than 900 effective samples were retained to build the corresponding directed prototype graph.} 
%润色后的
Standard nearest-neighbor retrieval based solely on semantic similarity is insufficient for this task, as the objective is to identify targets that are plausible under a specific transformation relation rather than those that merely resemble the input text. Furthermore, the graph is inherently multi-relational: distinct transformation types encode specific semantic directions that would be collapsed by a relation-agnostic model. Consequently, we compute semantic embeddings for all fragment nodes as initial features and employ a Relational Graph Convolutional Network (R-GCN). The R-GCN is trained over the multi-relational edge structure to learn node representations that capture transition regularities conditioned on relation type, transcending surface textual similarity. To rank candidate targets, we utilize a DistMult scoring function: 
%bq版本
% We did not adopt a simple nearest-neighbor retrieval based solely on semantic similarity, nor a graph model that ignores relation types. The reason is that \textit{Transform} is not merely a task of finding similar expressions, but a relation-conditioned transformation task: the goal is to retrieve target expressions that are plausible under a specific transformation type, rather than fragments that only resemble the input text on the surface. Since different shift types correspond to different semantic transformation directions, ignoring relation types would collapse these distinctions. We therefore use R-GCN to model relations under different shift types and learn relation-constrained transformation patterns. Semantic embeddings are computed for all fragment nodes as initial features, and a \emph{Relational Graph Convolutional Network} (R-GCN) is trained over the multi-relational edge structure to learn node representations that capture transition regularities conditioned on relation type rather than surface textual similarity. To rank candidate targets, we adopt a DistMult scoring function:
\begin{equation}
f(s,r,t)=\sum_{k=1}^{d} H_{s,k}\cdot R_{r,k}\cdot H_{t,k}
\end{equation}
%润色后的
where $H$ denotes the R-GCN output representation matrix, $R_r$ is the learned relation vector for transformation type $r$, and $d$ represents the embedding dimension. The $K$ highest-scoring candidates are then utilized as structural rewriting prototypes (see Appendix~\ref{apx:graphbuilder_algorithm} for detailed construction and training algorithms).
%bq版本
% where $H$ is the R-GCN output representation matrix, $R_r$ is the learned relation vector for shift type $r$, and $d$ is the embedding dimension. The top-$K$ highest-scoring candidates serve as \emph{structural rewriting prototypes} \textcolor{black}{(graph construction and training details are provided in Appendix~\ref{apx:graphbuilder_algorithm})}.

\paragraph{Online Retrieval and Constrained Rewriting.}
%润色版本
At runtime, when a user invokes the \opr{Transform} operation on a \WHV{} fragment $f$, the system identifies its specific role and current abstraction level. It then predicts the most contextually appropriate transformation type based on the fragment's textual content and its node context. 
Utilizing the learned scoring function, the system retrieves the top-$K$ transition prototypes from the corresponding role-specific subgraph. Rather than surfacing these prototypes directly to the user, the system injects them into the LLM prompt as structural constraints. This mechanism mandates that the generated rewrite: (1)~preserves the original abstraction level and \WHV{} role, and (2)~aligns with the semantic direction exemplified by the retrieved transitions. The resulting output is a perspective-shifted re-expression grounded in observed design reasoning patterns—structurally stable, interpretable, and ready for immediate re-entry into subsequent cluster, \opr{Merge}, and \opr{Steering} operations (see Appendix~\ref{apx:prompt-rgcn-rewriting} for full prompt templates).
% At runtime, when a user invokes \opr{Transform} on a \WHV{} fragment $f$, the system identifies its \WHV{} role and current abstraction level, predicts the most contextually appropriate shift type from the fragment text and its node context, and retrieves the top-$K$ transition prototypes from the corresponding role-specific subgraph via the learned scoring function. These prototypes are not surfaced directly to the user; instead, they are injected into the LLM prompt as \emph{structural constraints}, requiring the rewrite to (1)~preserve the original abstraction level and \WHV{} role, and (2)~move along the semantic direction exemplified by the retrieved transitions. The result is a perspective-shifted re-expression grounded in observed design reasoning patterns—structurally stable, interpretable, and directly re-enterable into subsequent \opr{Analyze}, \opr{Merge}, and \opr{Steering} operations (full prompts in Appendix~\ref{apx:prompt-rgcn-rewriting}).

\subsubsection{Stage 3: Cross-dimensional Recombination (DG3)}\label{sec:pip_S3} %✅
% To support structured cross-dimensional recombination (I3), we implement merge operators that systematically combine \WHV{} fragments according to the patterns observed in Study 2 (I3 in \S\ref{sec:fs-study2-findings}), including two approaches in various LLM prompts: 
%     \textit{(I) flexible recombination} when manipulating \WHV{} fragments (\texttt{Op\_VW}, \texttt{Op\_HV}, \texttt{Op\_WH}, or \texttt{Op\_WHV}) in multiple prompt settings and 
%     \textit{(II) four fixed recurrent modes} when manipulating the whole nodes, including 
% \texttt{Brainstorm} (75\% How+100\% Value), 
% \texttt{Create Experimental Innovation} (75\% How+50-75\% Value), \texttt{Create Future Vision} (100\% How+Value), 
% \texttt{Create Product Concept} (50\% How+50\% How) (full prompts in Appendix \ref{apx:prompt-fourmerge}). 

To support structured cross-dimensional recombination (DG3), we implement a set of merge operators to systematically combine nodes or \WHV{} fragments (Figure \ref{fig:pipeline}D). For whole nodes, the system identifies the \WHV{} attributes of the raw nodes via an LLM to apply different operators %---\texttt{Op\_VW}, \texttt{Op\_HV}, \texttt{Op\_WH}, and \texttt{Op\_WHV}---
 to support structured recombination across different dimensions. For \WHV{} fragments, the system provides four predefined targeted direction found in Study 2, DG3 \footnote{In Study 2, we identified recurring patterns where participants utilized typed fragments at varying levels of abstraction ($L1$–$L4$) for different purposes. Specifically, these were characterized as follows: \texttt{Brainstorm} (L3-How + L4-Value), \texttt{Experimental Innovation} (L3-How + L3--L4 Value), \texttt{Future Vision} (L4-How + L4-Value), and \texttt{Product Concept} (L2-How + L2-How).}. %including \texttt{Brainstorm} (75\% How + 100\% Value), \texttt{Create Experimental Innovation} (75\% How + 50--75\% Value), \texttt{Create Future Vision} (100\% How + Value), and \texttt{Create Product Concept} (50\% How + 50\% How). 
 For \WHV{} fragments that do not match any predefined direction, the system generates the final merged result using a default merge mode (details in Appendix~\ref{apx:prompt-merge}).

\subsubsection{Stage 4: Multi-Layered Representations (DG1, DG3)}\label{sec:system_designspacegraph} 

\begin{figure}[h]
    \centering
    \includegraphics[width=\linewidth]{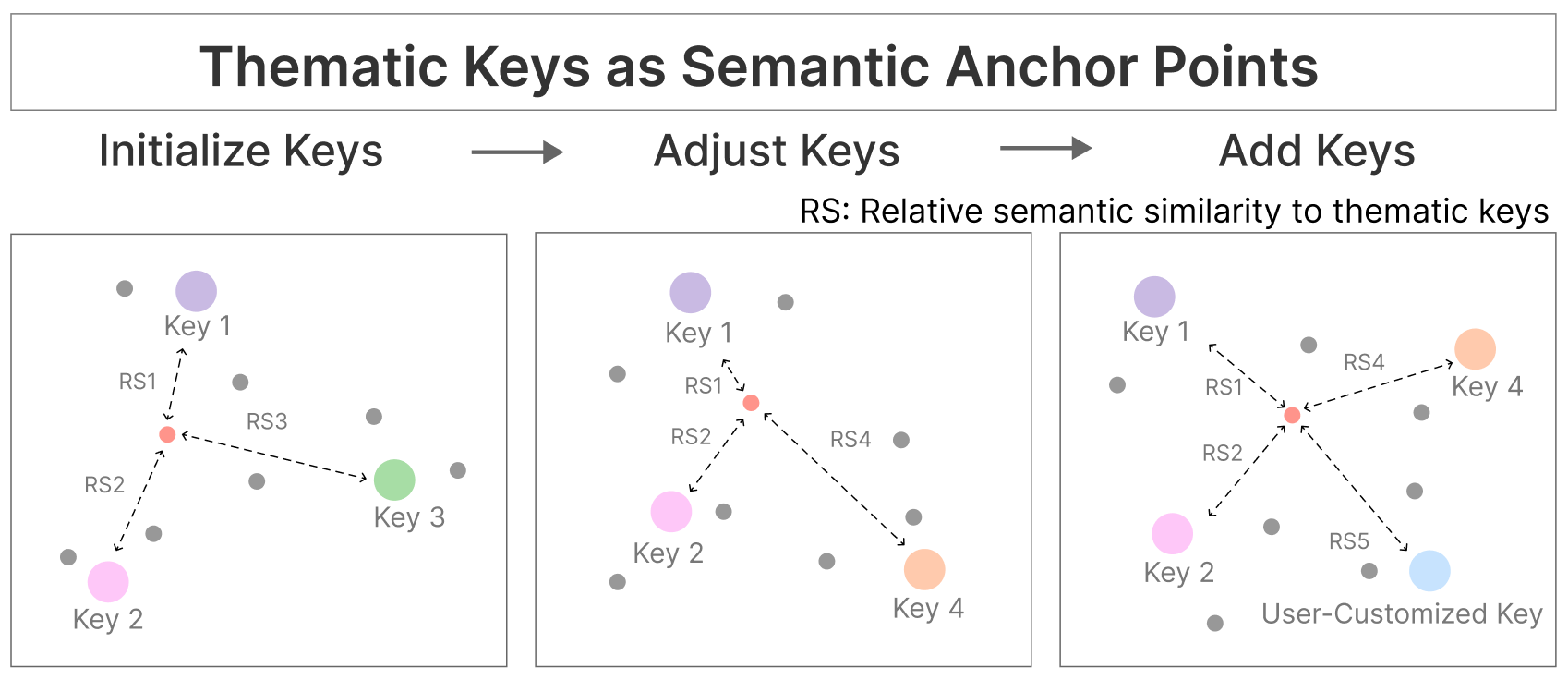}
    \caption{Thematic-key-based semantic organization in the canvas. 
    User can flexible organize nodes using these thematic keys. 
    }
    \label{fig:cluster_graphicalapproches}
\end{figure}

To make relational structure spatially perceptible (DG1), the pipeline positions nodes in a canvas layout determined by their similarity to a set of user-defined thematic \emph{keys} (termed Theme Tag in \S\ref{sec:system_multiview}) (Figure \ref{fig:pipeline}E). These thematic keys serve as semantic anchor points that function as interpretable reference axes in the canvas. The system requires a minimum of three keys, inducing a multi-dimensional semantic layout in which each node's position reflects its joint similarity profile across all reference axes; adding further keys progressively reveals cross-cutting structure among fragments (Figure~\ref{fig:cluster_graphicalapproches}). Candidate keys can be introduced either automatically, by applying KMeans to node embeddings followed by LLM-generated labels, or manually by the user, thereby supporting both exploratory discovery and goal-directed structuring of the design space %(implementation details in Appendix~\ref{apx:clustermode})
.

\subsection{\N{} Interaction Design}
\begin{figure*}
    \centering
    \includegraphics[width=\linewidth]{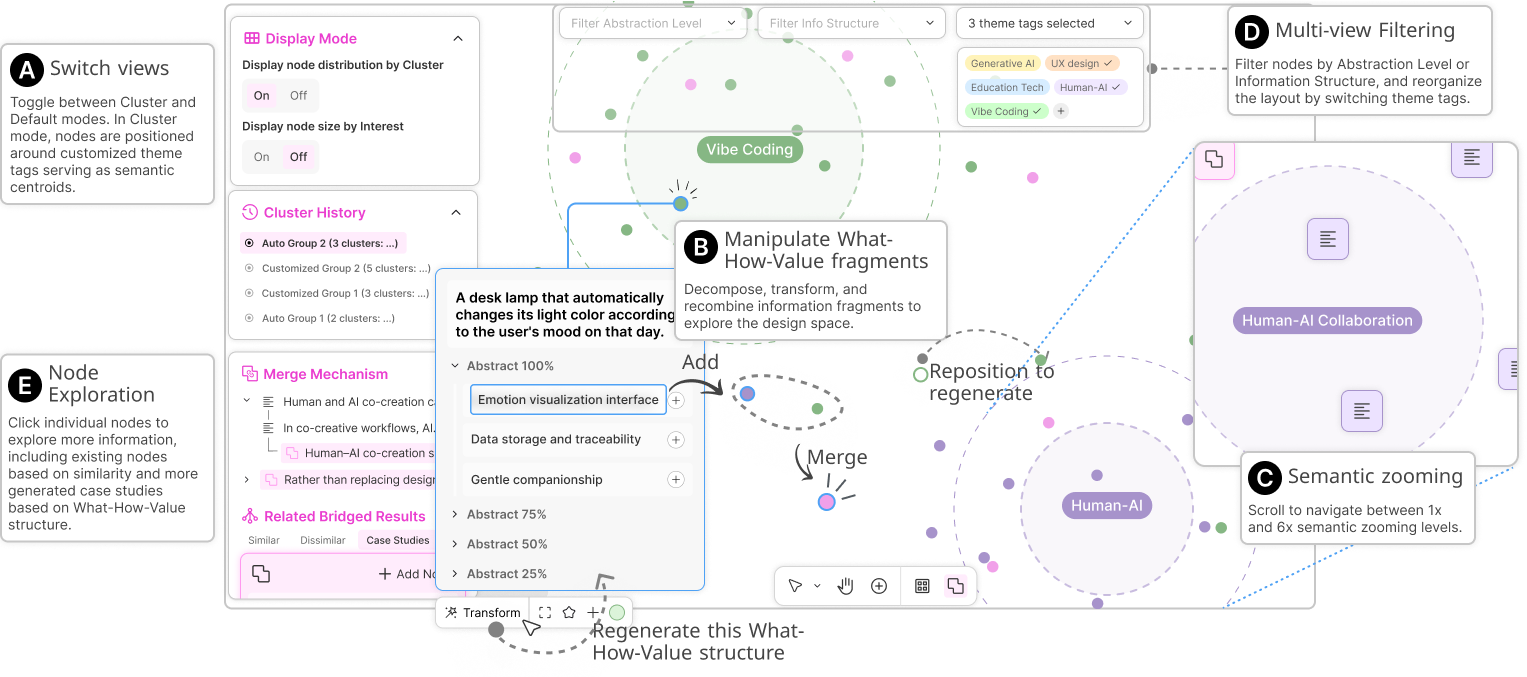}
    \caption{\N{} interface, a graph-based LLM system that supports creative exploration through structured, manipulable units. 
    (A) \textbf{Switch Views} %Toggling between Cluster and Default modes to organize nodes by semantic centroids. 
    (B) \textbf{Manipulate Fragments} %Decomposing, transforming, and recombining \WHV{} fragments to explore the design space. 
    (C) \textbf{Semantic Zooming} %Navigating the view across 1x--6x magnification levels. 
    (D) \textbf{Multi-view Filtering} %Filtering nodes by abstraction levels or information structures. 
    (E) \textbf{Node Exploration} %Accessing detailed node content and its \WHV{} decomposition through direct interaction.
    }
    \label{fig:teaser}
\end{figure*}
\N{} presents as an AI-assisted creativity tool built on a diagramming canvas (Figure \ref{fig:teaser}). Each piece of inspiration is represented as a node containing \WHV{} manipulable units decomposed from its semantic content. The following interaction mechanisms operationalize DG1--DG3.

\subsubsection{Graphical Structure: \WHV{} with L1--L4 as Manipulable Units (DG1, DG2)} \label{sec:system_interaction+abstraction}
The system automatically parses each node into \WHV{} manipulable units—each a short sentence expressing a coherent meaning within one functional role. 
% \N{} presents as a AI-assisted creativity tool with diagramming and visualizing approaches. In this system, each information presents as a node, and contains several \WHV{} fragments decomposed from semantic meaning.  

\textbf{Interaction: Analyze and Transform.}
Each node supports two core operations (Appendix, Figure~\ref{fig:analyze}): \opr{Analyze} parses the node content into \WHV{} fragments, making implicit structure explicit; \opr{Transform} applies an abstraction operator (\texttt{Op\_ELEVATE}, \texttt{Op\_MECH}, or \texttt{Op\_VALUE}) to perform a controlled perspective shift and re-expression on each \WHV{} fragment while keeping its abstraction level unchanged, thereby supporting the perspective transformation emphasized in DG2.

\textbf{Abstraction Levels: L1--L4.}
    These fragments are divided into four abstraction level, shown as \opr{Abstraction L4: 100\%--L1: 25\%} (Appendix, Figure \ref{fig:analyze}). These semantic fragments structure how raw information is decomposed into manipulated units.

Users can drag any analyzed or transformed fragment out to create a new node on the canvas, which itself supports further \opr{Analyze} and \opr{Transform}. These operations externalize the tacit decomposition and abstraction processes identified in Study 2, corresponding to Stage 1 (\S\ref{sec:pip_S1}) and Stage 2 (\S\ref{sec:pip_S2}). 
% \begin{figure}
%     \centering
%     \includegraphics[width=0.8\linewidth]{Figures/analyze_transform.png}
%     \caption{User flow of Analyze and Transform: Users click Analyze to parse \WHV{} fragments (A), then Transform to shift perspective to regenerate \WHV{} fragments (B). A fragment can be dragged to create a new node (C), which also supports further parsing (D).}
%     \label{fig:analyze}
% \end{figure}
~
% To achieve \textbf{DG1,} \N{} system incorporates graphical and visualizing approaches to manipulate information. 
% \subsubsection{Graphical Structure for \WHV{} as Manipulable Units} 
% Our system automatically breaks each information into manipulable units, each defined as one short sentences that express a coherent meaning from What, How, and Value. 
~
% \paragraph{Interaction} Each node and unit is represented as an interactive manipulable units where enables \opr{Analyze} and \opr{Transform} from one piece information into \WHV{} fragments (Figure~\ref{fig:teaser}). 
~
%     These fragments are divided into four abstraction level, shown as ``\opr{Abstraction 100\%--25\%}''(details in Section \mbox{\ref{sec:system-Abstraction}}). \mbox{\aq{one example}} These semantic roles structure how raw information is externalized and manipulated in the canvas.
~
% User click \opr{Analyze} option below one node...; Once one node is analyzed, \opr{Transform} option emerges, enabling to generate alternative perspectives of \WHV{} fragments across four abstraction level (Figure \mbox{\ref{fig:analyze}}).  
User can also explore related content: selecting a node triggers the side panel to display semantically \emph{similar} and \emph{dissimilar} nodes as well as related \emph{case studies} (Figure \ref{fig:teaser} E), enabling cross-domain inspiration. %and supporting the parallel thinking identified 
% in Study 2. 

\subsubsection{Cross-dimensional Recombination (DG3)} 
This interaction operationalizes Stage 3's cross-dimensional recombination (\S\ref{sec:pip_S3}). Users select two fragments or nodes on the node or canvas and trigger \opr{Merge} (Appendix, Figure~\ref{fig:merge_userflow}). The system identifies the \WHV{} roles of the selected fragments and applies the corresponding operator (\S\ref{sec:pip_S3}). %Multiple inspiration directions are generated and diversity-scored; the top-$k$ non-redundant results are displayed in the side panel for inspection and selection.  

\subsubsection{Multi-View Spatial Organization (DG1)}\label{sec:system_multiview}
This helps users perceive associations among fragments, 
% (Appendix, Figure~\ref{fig:2displymode}), 
corresponding to \S\ref{sec:system_designspacegraph}: 
    \textbf{Default Mode} 
    supports free spatial organization by the user, enabling manual grouping and proximity-based structuring. 
    \textbf{Cluster Mode} 
    positions nodes in a shared high-dimensional embedding space, where theme tags function as semantic centroids and nodes are arranged according to centroid-relative similarity %(details in Appendix~\ref{apx:clustermode})
    . When users introduce additional theme tags, the system generates new semantic centroids, triggering a reallocation of nodes based on updated similarity relationships. Interface and users flow is shown in Appendix, Figure~\ref{fig:2displymode} and Figure~\ref{fig:clusterview}. 
    
% \begin{figure}[h]
%     \centering
%     \includegraphics[height=3.8cm]{Figures/display_mode.png}
%     \caption{Two display modes: cluster (A) and default (B).}
%     \label{fig:2displymode}
% \end{figure}

    % represents as a shared high-dimensional embedding space, \aq{what is position distribution of themes?}where themes function as semantic centroids and nodes are positioned according to centroid-relative similarity (details in Appendix \ref{apx:clustermode}) User flow are shown in Figure~\ref{fig:cluster_userflow}. 
    % Spatial organization is defined by a dynamically evolving set of theme centroids. When users introduce additional themes (Figure \ref{fig:teaser}\aq{/}), the system generates new semantic centroids within the shared embedding space, triggering a reallocation of nodes based on updated similarity relationships. 
~

    \textbf{Node~Size by Interest}: 
    The system visually encodes node size by interaction frequency (Appendix, Figure~\ref{fig:display_interest}), counting \opr{Analyze}, \opr{Transform}, fragment drag-out, \opr{Merge}, Steering, and click operations. This provides users with a spatial map of their own engagement, supporting reflective awareness of unexplored areas. 
    % This system also supports visual encode of node size based on interesting (Figure~\ref{fig:display_interest}), counting number of operations, including \opr{Analyze}, \opr{Transform}, dragging out \WHV{} fragment to a node, and \opr{Merge} \aq{Have to add ``Steering'' and ``Clicking'' to check}. 
% \begin{figure}[h]
%     \centering
%     \begin{subfigure}{\linewidth}
%         \centering
%         \includegraphics[width=\linewidth]%[height=3.8cm]
%         {Figures/cluster_userflow.png}
%         % \caption{}
%         \label{fig:cluster_userflow}
%     \end{subfigure}
%     \hfill
%     % \begin{subfigure}{0.44\linewidth}
%     %     \centering
%     %     \includegraphics[height=3.8cm]{Figures/display_interest.png}
%     %     \caption{}
%     %     \label{fig:display_interest}
%     % \end{subfigure}
%     \caption{%(a) User flow of clustering with (b) visual encoding node size by interest level. \aq{detail ABCDE in (a)} These theme categories are visualized in distinct color hues and spatial clustering, enabling users to quickly identify and reorganize design elements. \aq{detail AB in (b)}
%     %(a) 
%     User flow of Cluster Mode: theme tags function as semantic centroids, and nodes are repositioned based on similarity. %(b) Node size encoded by interaction frequency—larger nodes indicate higher users engagement.
%     }
%     \label{fig:clusterview}
% \end{figure}

\subsubsection{Navigation and Filtering (DG1, DG2)} 
To support parallel thinking and multi-level exploration, \N{} provides the following navigation (Figure~\ref{fig:teaser} C, D): 
% \begin{itemize}
%     \item \textbf{Semantic zooming (1$\times$--6$\times$)} \aq{have to consider} (Figure \ref{fig:zoomlevel})
%     \item \textbf{Filtering by abstraction level (L4: 100\%--L1: 25\%):} Users can filter nodes according to abstraction level built from either raw information or \WHV{} fragments to focus on the most relevant nodes (Figure \ref{fig:teaser}). This allows reducing visual clutter while retaining contextual information. 
%     \item \textbf{Filtering by \WHV{} type:} Users can filter nodes according to \WHV{} fragment types (i.e., What, How, or Value). The source of nodes can be either raw information or existing \WHV{} fragments. This enables exploration of specific abstraction or reasoning layers. 
% \end{itemize}

\begin{figure*}
    \centering
    \includegraphics[width=\linewidth]{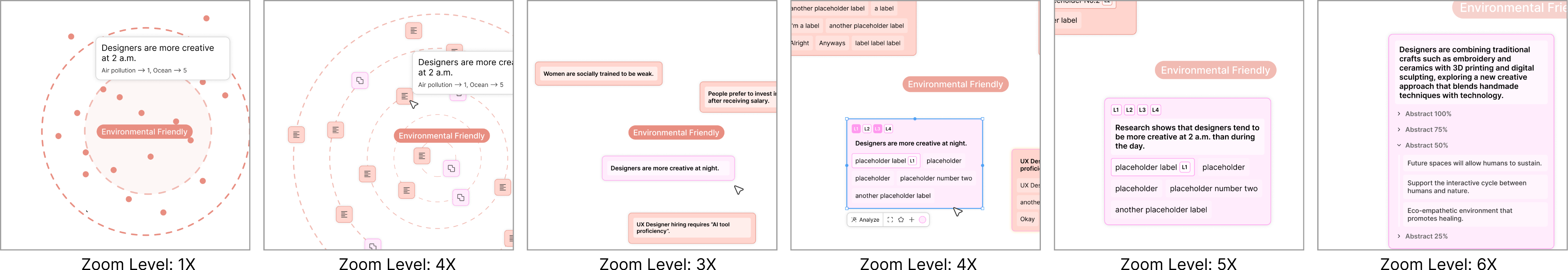}
    \caption{Semantic zooming (1$\times$--6$\times$): information granularity and interface layout adapt dynamically to zoom level, supporting cognitive scalability from high-level structure to detailed fragment content.}
    \label{fig:zoomlevel}
\end{figure*}

\begin{itemize}
    \item \textbf{Semantic zooming (1$\times$--6$\times$):} Information granularity and layout adapt dynamically to zoom level, allowing users to move fluidly between high-level structure and detailed fragment content (Figure~\ref{fig:zoomlevel}).
    \item \textbf{Filtering by abstraction level (L1--L4):} Users filter nodes by abstraction level to focus on fragments at a target specificity—e.g., showing only L3 (Principle) fragments to identify transferable design values, or only L1 (Fact) fragments to stay grounded in concrete observations.
    \item \textbf{Filtering by \WHV{} type:} Users filter nodes by functional role (What, How, or Value) to explore specific reasoning dimensions independently.
\end{itemize}

%缩短版
\subsubsection{Navigation and Steering Modes (DG1, DG2)} To balance free exploration with structured control, \N{} offers two node interaction modes (Figure~\ref{fig:teaser} B): \textbf{Navigation Mode:} Users reposition nodes in Default layouts to support spatial reorganization and externalization (DG1). \textbf{Steering Mode:} Dragging nodes in Cluster layout triggers theme-constrained rewriting. The system computes a proximity-weighted signal over $K$ nearest theme centroids, modulated by temperature $\tau$, to regenerate content toward the target direction. %This operationalizes Perspective Transformation (DG2): a spatial drag becomes a controlled semantic shift, re-estimating the node’s position via cosine similarity to ensure layout-content consistency. 
This enables Perspective Transformation (DG2): a spatial drag becomes a controlled semantic shift, aligning layout with content via cosine similarity. 

\subsection{System Implementation}
NexusAI is a web-based ideation environment utilizing a Vue frontend%\footnote{\url{https://vuejs.org/}}
and a Spring Boot backend.\footnote{\url{https://spring.io/projects/spring-boot}} We leverage Qwen-Plus through the DashScope API provided by Alibaba Cloud Model Studio\footnote{\url{https://www.alibabacloud.com/help/en/model-studio/qwen-api-via-dashscope}.} for LLM-based generation and reasoning, alongside DashScope text embeddings (specifically \texttt{text-embedding-v4})\footnote{\url{https://www.alibabacloud.com/help/en/model-studio/embedding}.} for semantic retrieval and clustering.The frontend provides a node-based visual canvas, while the backend manages state and persistence through RESTful APIs.\footnote{\url{https://restfulapi.net/}.} Crucially, instead of plain text, ideas are stored as structured objects via a 12-slot representation; each \WHV{} fragment is indexed by its abstraction level, functional role (What/How/Value), ordinal position, and summary title. This persistent schema enables the systematic externalization and navigation of fragmented design information.

\section{User Study}
    To investigate how the proposed \N{} facilitates design space exploration, we conducted a within-subject study comparing two systems: a conventional dialogue-based LLM interface versus our structured interaction CA pipeline in \N{}. 
%v2-15Feb2026%
The study investigated the following questions: 
\begin{enumerate}
    \item [\textbf{RQ1}]  How does the CA pipeline’s structured navigation facilitate exploration of design spaces? 
    \item [\textbf{RQ2}]  How does the CA pipeline’s granular manipulation mitigate design fixation during AI-mediated ideation? 
\end{enumerate}
% \begin{enumerate}
%     \item [\textbf{RQ1}] How do users interact with \N{} throughout the workflow of exploring a design space? 
%     \item [\textbf{RQ2}] How do users decompose and recombine ideas into structured manipulable units during exploration (DG1)? 
%     \item [\textbf{RQ3}] How abstraction shifting and semantic zooming influence the breadth and diversity of explored design space (DG2)? 
%     \item [\textbf{RQ4}] How various representations support cross-dimensional recombination and relational sense-making (DG3)? 
% \end{enumerate}

% The alignment between designers’ evolving exploration intentions and AI-generated structural transformations.

% Through this study, we aim to understand whether structuring LLM outputs as interactive artifacts can mitigate premature convergence, and scaffold non-linear, multi-perspective design space exploration.

%v1-before15Feb2026%
% Our research questions are: 

% \begin{itemize}
%     \item How can \N{} be used in creative workflows?
%     \item How does \N{} help with divergent exploration?
%     \item How does \N{} help with help with developing an design space?
%     \item What is the perceived utility of \N{}'s features?
% \end{itemize}
\subsection{Participants} 
Fourteen participants (Appendix \ref{apx:us-participants}, Table \ref{tab:us-participants}; 10 female, 4 male; aged 24-33, $M$=27.1, $SD$=2.66) were recruited via purposeful sampling through practitioner communities. Inclusion criteria required (1) at least one year of LLM use and (2) a minimum weekly LLM usage frequency. Participants were screened via a pre-study survey to ensure diverse creative backgrounds  (writing, conceptual thinking, design, and art) and to mitigate expertise-related variance.
~
The pre-study survey collected: (1) demographic information (age, gender, profession related to creativity/design), (2) creativity and design experience (duration, field of expertise), and (3) experience using generative AI tools (types, use cases). Participation was voluntary and uncompensated. Informed consent was obtained prior to the study. This study was approved by university IRB. 

\subsection{Study Design} %We compared our system to a baseline system in a within-subject study design. 
\subsubsection{Tasks}
We designed two parallel creative exploration tasks of equivalent cognitive demand, following counterbalanced assignment across conditions: 
% \begin{itemize}
    % \item 
    %\aq{two experts compare; counterbalanced}
    \emph{Task A:} Design a future \emph{personal focus assistant}—a system that helps individuals manage attention, cognitive states, and working rhythms in an era of constant digital interruption. 
    % \item 
    %\aq{to-be-update}
    \emph{Task B:} Design a future \emph{community memory archive}—a system that helps neighborhoods or social groups capture, preserve, and meaningfully revisit collective experiences and local knowledge.
% \end{itemize}

Both tasks were intentionally open-ended and domain-agnostic, incorporating the keywords \textit{``future''} and \textit{``design''} to encourage imagination beyond current technological constraints~\cite{shen_ideationweb_2025}. 
Tasks were piloted and validated by two HCI and design researchers to ensure they offered comparable creative degrees of freedom and cognitive complexity. 
% Tasks were piloted and confirmed to be of comparable difficulty. 
Each task was completed within 20 minutes.

\subsubsection{Baseline Rationale}~\label{sec:baseline}
%图形界面+对话界面
To isolate the effect of \mbox{\CA{}} pipeline, we designed a baseline system that matched \mbox{\N{}} in LLM capability and diagramming affordances (Appendix \ref{apx:baseline}, Figure~\mbox{\ref{fig:baseline}}). %, differing only in how \textbf{graphic structure} and \textbf{manipulable unit} were externalized and interacted. 
%保留什么功能-怎么和system一致
Both conditions used the same underlying LLM, identical prompt templates, and the same turn-based generation mechanism. %In both systems, AI outputs were generated only upon explicit user invocation, and no additional reasoning depth, hidden chain-of-thought prompting, or intermediate inference steps were introduced in either condition. 
The baseline provided an integrated node-based diagramming canvas with basic AI node generation, allowing participants to incrementally develop design space within a unified workspace. 
%去除什么-怎么和system不一致
In contrast, \mbox{\N{}} enabled users to explicitly curate and reorganize relationships among reasoning on nodes, which could then be referenced as structured context during subsequent generations. Importantly, %this structure did not introduce new generative functions or increase model capability; rather, it 
\N{} only altered how existing content was selected, composed, and interpreted for a exploring design space.

\subsubsection{Procedure}
% We conducted a counterbalanced within-subjects study design. 
Participants began with completing informed consent and receiving an introduction of goals and procedure. Task order and condition assignments were balanced within two groups: seven participants completed Task A with \N{} and Task B with the baseline; the remaining seven completed Task A with the baseline and Task B with \N{}. 
Each session lasted approximately 90 minutes and followed this sequence: 
\emph{(1) Introduction and demonstration (5 mins):} The experimenter introduced \N{}'s core features, using a brief walkthrough; 
\emph{(2) Free exploration (5 mins):} Participants explored \N{} freely with a warm-up prompt to build familiarity before the formal tasks; 
\emph{(3) two task sessions (20 mins each for each condition)}; 
\emph{(4) post-task feedback} questionnaires (10 min; detailed in \S\ref{sec:measurements}) and a final \emph{semi-structured interview} (15-25 min; see Appendix \ref{apx:interviewguide}) was conducted after both tasks were complete. 
During task sessions, think-aloud protocol~\cite{Ericsson1980VerbalRA} was encouraged throughout. All sessions were screen and audio recorded.

\subsubsection{Measurements}
\label{sec:measurements}
We evaluated both systems across three aspects: System Usability, Creativity Support and Cognitive Load. System usability was evaluated by a 7-point 
\textit{Self-defined Likert scale items} (Appendix \ref{apx:self-definedlikertscale}) across perceived structuring design space exploration, ease of decomposing and recombining ideas, and effectiveness of multi-level abstraction. System Usability was also evaluated by the \textit{System Usability Scale (SUS)} assessing overall system usability, yielding a single score (0--100). Scores above 68 indicate above-average usability.  
~
Creativity Support was measured by CSI~\cite{cherry_quantifying_2014} across six dimensions (Exploration, Expressiveness, Immersion, Enjoyment, Results Worth Effort, and Collaboration). 
~
Cognitive Load was measured by NASA-TLX covering Mental Demand, Physical Demand, Temporal Demand, Performance, Effort, and Frustration.  
Both systems logged various types of time-stamped events based on participants’ interactions during the study, including higher-level ideation actions, interface organization and control actions, supporting exploration actions, and lower-level navigation behaviors, as well as written prompts and AI responses. 

\subsubsection{Data Analysis}
\label{sec:dataanalysis}

% \paragraph{Quantitative analysis.}
%只用Wilcoxon的版本
% To evaluate the effectiveness of \N{}, we synthesized subjective, qualitative, and behavioral data. 
For all subjective and structural metrics, we performed \textbf{Wilcoxon signed-rank tests} to compare paired scores between \N{} and the baseline. This non-parametric approach was selected due to the within-subject design and the non-normal distribution of our ordinal data. For each test, we report the level of significance ($p$) and the effect size ($r$) to characterize the magnitude of the observed effects.For qualitative data, including post-study interview transcripts and think-aloud recordings, we conducted thematic analysis~\cite{braun2006thematic}. Two researchers independently coded the data according to the design principles, resolving discrepancies through iterative discussion until consensus was reached. 

For behavioral log analysis, we reconstructed exploration graphs to quantitatively characterize user trajectories. These were analyzed across two dimensions:
For RQ1, we computed average root-to-leaf depth, and re-engagement rates to measure the continuity and persistence of exploration. Task alignment was further assessed via Coverage and Coherence-to-Prompt.
For RQ2, we analyzed average out-degree, and maximum layer width to quantify the breadth and scale of the design space expansion. 
To facilitate cross-session comparisons, all interaction timestamps were normalized to a $[0, 1]$ scale, allowing for both individual timeline visualization and aggregate temporal trend analysis. 

\section{Findings}
    % \subsection{Design Space Exploration and Divergence}

% \subsection{Structured Sensemaking through Spatialized Units}

% \subsection{Vertical Movement across Abstraction Levels}

% \subsection{Cross-Domain Integration and Transferable Insights}

% \subsection{Idea Evolution and Reusable Design Trajectories}

% \begin{figure*}[t]
%     \centering
%     \includegraphics[width=0.8\linewidth]{Figures/Likert.png}
%     \caption{Comparison between \N{} (left) and baseline (right) with a 7-point Likert scale.}
%     \label{fig:likertchart}
% \end{figure*}

%1Apri-Finding Structure
% https://gemini.google.com/share/804ff182a343
\begin{figure*}[htbp]
    \centering
    % 第一个子图 (Analyze and Transform)
    \begin{subfigure}[b]{0.47\linewidth}
        \centering
        \includegraphics[height=3.8cm]{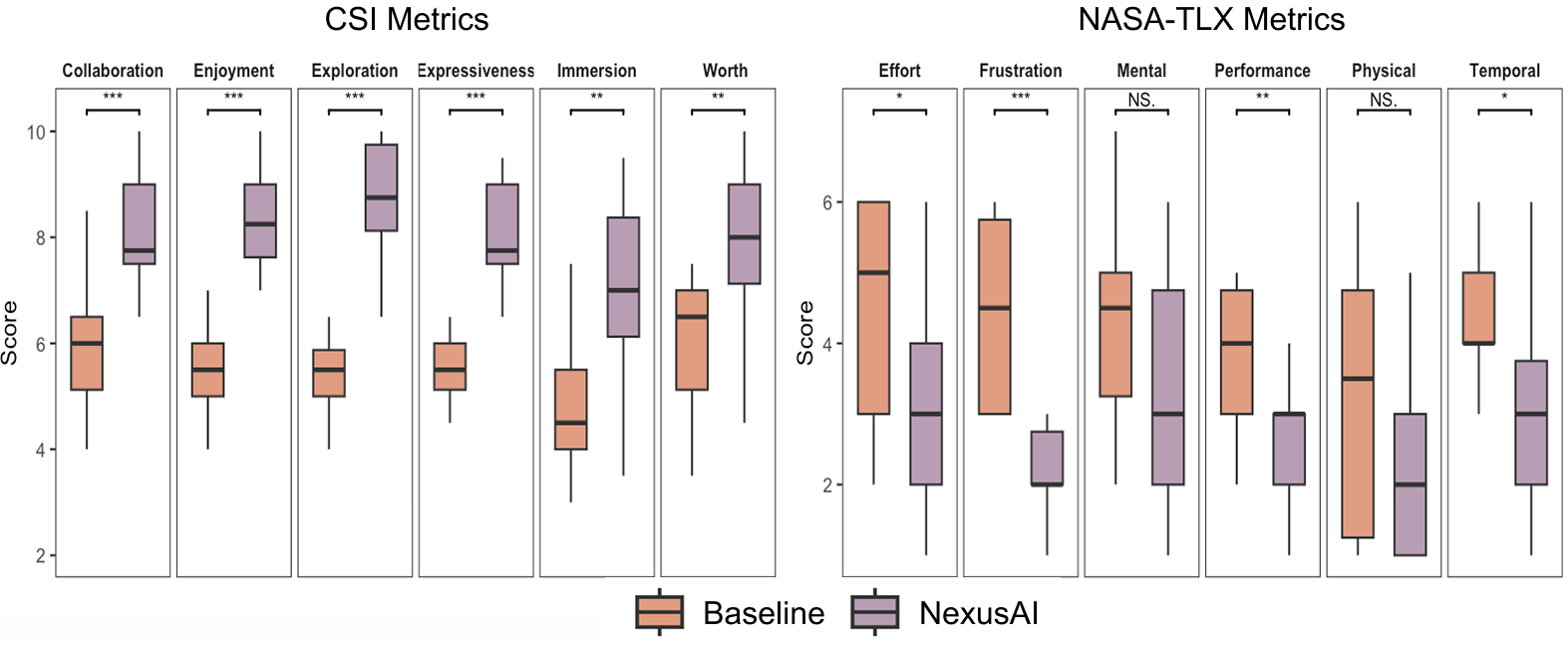}
        \caption{}
        \label{fig:boxplot}
    \end{subfigure}
    \hfill
    % 第二个子图 (Merge)
    \begin{subfigure}[b]{0.49\linewidth}
    \centering
    \includegraphics[height=3.8cm]{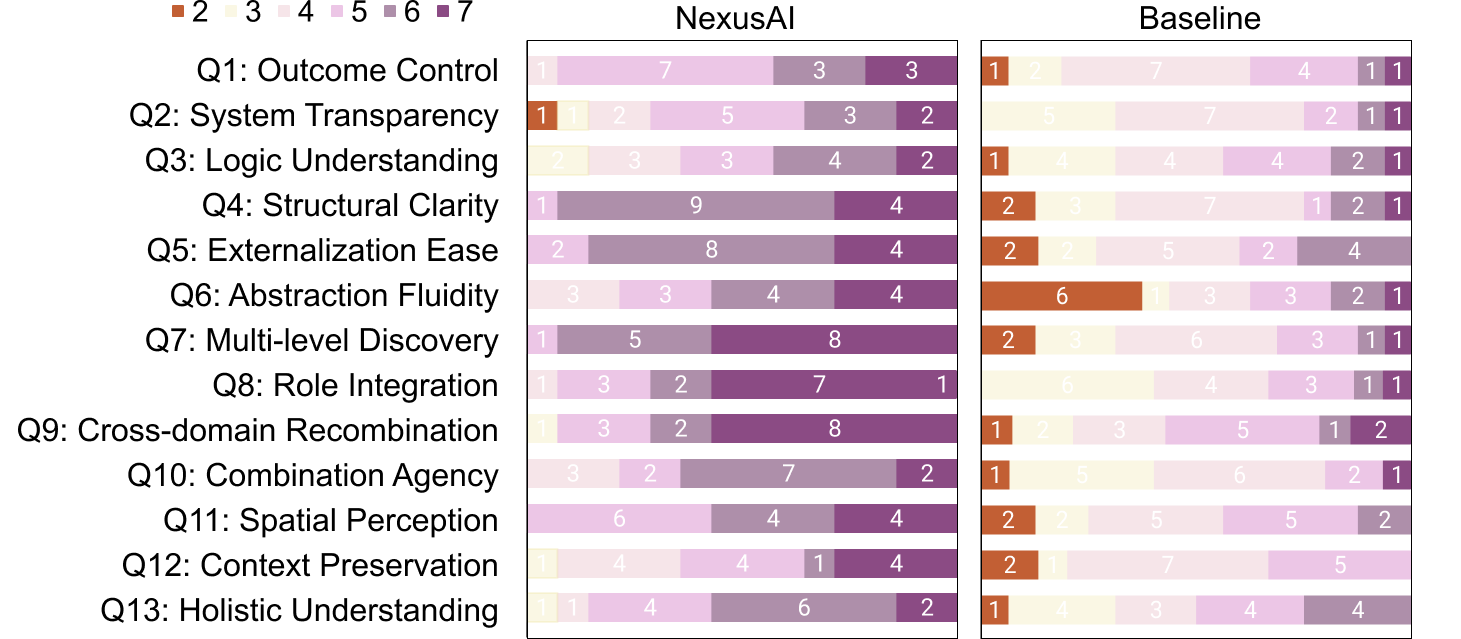}
    \caption{}
    \label{fig:likertchart}
    \end{subfigure}

    \caption{Comparison between \N{} and baseline with (a) CSI metrics, NASA-TLX metrics and (b) a 7-point Likert scale.}
    \label{fig:datachart}
\end{figure*}

% We report findings from participants' survey responses, think-aloud protocols, post-study interviews, and system interaction logs, organized around four aspects aligned with our design goals.

%------------------------------------------------------------
\subsection{NexusAI Shifts Exploration from Reactive Generation to Structured, Intent-Driven Progression (RQ1)}
\label{sec:findings-workflow}
The average system usability scores computed from SUS were significantly greater for our system, compared to the baseline ($M_{Nx}$=69.3 vs. $M_{baseline}$=57.3, $p$=0.011*). 
The implementation of the CA pipeline fundamentally restructured user exploration from superficial sequences into persistent, intent-driven trajectories. Quantitative analysis of system logs confirms that \N{} facilitates a more purposeful exploration compared to the baseline. Users in the \N{} condition developed significantly more complex evolutionary structures, evidenced by 
% a higher number of explicit derivation edges 
%     ($M_{Nx}=69.57$ vs. $M_{baseline}=2.31$) and 
a deeper average root-to-leaf exploration path 
    ($M_{Nx}=1.01$ vs. $M_{baseline}=0.47$, $p$=0.002**, $r$=0.839). 
This sustained focus is further supported by a markedly higher re-engagement rate with core skeleton nodes 
    ($M_{Nx}=0.367$ vs. $M_{baseline}=0.097$, $p$=0.008**, $r$=0.688), 
indicating that users strategically refined existing ideas rather than abandoning them. \textcolor{black}{Critically, this exploration was not random but highly task-aligned, as evidenced by a significantly higher Coverage ($M_{Nx}=0.407$ vs. $M_{baseline}=0.079$, $p=0.004^{**}$) and CoherenceToPrompt ($M_{Nx}=0.274$ vs. $M_{baseline}=0.053$, $p=0.009^{**}$).} These metrics %, %combined with superior scores for perceived steering capability ($M_{Nx}=5.57$ vs. $M_{baseline}=4.31$, $p=0.0068^{**}$, $r=0.783$) and logic transparency ($M_{Nx}=5.07$ vs. $M_{baseline}=4.31$, $p=0.0474^{*}$, $r=0.598$), 
demonstrate that the framework effectively anchors deep exploration to core design requirements.

%------------------------------------------------------------
% \textbf{\N{} enabled a goal-directed exploration workflow that moves from the big picture to the details.} 
% Participants using \N{} consistently organized information into \WHV{} units before generating new directions, while baseline participants tended to generate content immediately and attempt to impose structure afterward. 
% % \aq{[INSERT: Mdn comparison of decomposition event timing; $p < .05$.]} 
% P3 explained, \textit{``I knew what I was working with from the beginning—the types were already there, so I wasn't just dumping ideas and hoping something stuck.''} P11 similarly noted that structured units helped them \textit{``see where the gaps were without re-reading everything.''} This pattern reduced exploratory overhead and kept participants oriented toward generative next steps rather than maintenance of context. 

\textbf{\N{} enabled a ``macro survey, local analysis, cross-domain recombination'' workflow that shifted users from passive responders to active directors of exploration.} Rather than turn-based interaction with AI, participants first used cluster and semantic-zoom views to survey the overall distribution of ideas, then drilled into individual nodes via \opr{Analyze} and \opr{Transform}, and finally composed outputs across themes via \opr{Merge}. 
% \aq{[INSERT: Mdn event-type distribution across normalized time; $p < .05$.]} 
Interaction events distribution reflects this description (Appendix, Figure \ref{fig:Nexus timeline}). 
Each participant' interaction is shown in Appendix, Figure \ref{fig:log}. 
This structured progression helped participants track their own reasoning trajectory: P8 noted the canvas \textit{``spreads information visually in front of me—at least I can tell where I am and roughly where I'm going,''} contrasting this with dialogue interaction where content \textit{``disappears as soon as I switch away.''}

\textbf{Participants reported substantially greater transparency into AI reasoning and control over its outputs.} Participants agreed significantly more that they could steer AI outputs in line with their intent (Figure \ref{fig:likertchart}) (Q$_1$: $M_{Nx}$=5.57 vs. $M_{baseline}$=4.31, $p$=0.007**, $r$=0.783), recognize the system's reasoning process (Q$_2$: $M_{Nx}$=5 vs. $M_{baseline}$=4.12, $p$=0.023*, $r$=0.678), and understand the logic behind its responses (Q$_3$: $M_{Nx}$ =5.07 vs. $M_{baseline}$=4.31, $p$=0.047*, $r$=0.598) with \N{} than with the baseline. P8 stated, \textit{``I could see exactly which fragments the AI was building from. That transparency made me trust the output more.''} 
P10 similarly contrasted the systems: \textit{``Compared to pure text-based AI, Nexus goes deeper. Through repeated Analyze, I can extract much deeper or more advanced ideas. GPT tends to stay on the surface; Nexus's fragmentation gives me more to think about.''} 
%Similarly, P10 stated, \textit{``Through continuous analysis, this system allowed me to extract a great many deeper or more advanced ideas''}. In contrast, participants frequently described baseline that is remaining superficial meaning and uncertainty about why the AI responded as it did, which they found disorienting during open-ended exploration.

%------------------------------------------------------------
\subsection{Granularity Manipulation Facilitating Multidimensional Divergence and Creative Synthesis (RQ2)}
% \N{} supports greater creativity from CSI score during such exploration ($M_{Nx}$=96.5 vs. $M_{baseline}$=66.1, $p$=0.0198**; Figure \ref{fig:boxplot}). 
The granular manipulation of idea fragments through the CA pipeline significantly expanded both the scale and structural variety of the design space. Log data reveals that \N{} provides superior support for structured divergent thinking, characterized by a significantly higher average out-degree per node ($M_{Nx}=0.98$ vs. $M_{baseline}=0.30$, $p$=0.004**, $r$=0.781)% \textcolor{black}{and a vastly expanded design space width ($Max\_width_{Nx}=66.93$ vs. $4.21$, $p$=0.001**, $r$=0.881)}
, which reflects a greater capacity for horizontal expansion. 
~
%In extreme cases, the system supported massive parallel exploration, with a maximum layer width reaching far beyond the baseline ($Max_{Nx}=138$ vs. $Max_{baseline}=8$). 
This increased divergence was achieved alongside a reduction in perceived workload ($M_{Nx}=3.26$ vs. $M_{baseline}=4.17$, $p=0.049$*) and temporal demand ($M_{Nx}=3.00$ vs. $M_{baseline}=4.43$, $p=0.049$*), allowing users to externalize inspirations with significantly greater ease ($M_{Nx}=6.14$ vs. $M_{baseline}=4.06$, $p=0.004$**, $r=0.828$). 
~
The effectiveness of these primitives in fostering novel connections is further validated by higher scores in cross-domain recombination ($M_{Nx}=6.14$ vs. $M_{baseline}=4.19, p=0.006^{**}, r=0.777$) and overall creative quality as measured by the CSI ($M_{Nx}=96.50$ vs. $M_{baseline}=66.10, p=0.020^{*}$; Figure \ref{fig:boxplot}).

\subsubsection{Decomposition Helps to Structure and Recombine Ideas Purposefully}
\label{sec:findings-dg1}
%------------------------------------------------------------
\textbf{Breaking nodes into units helped participants surface sub-idea structure invisible in undifferentiated text.} Participants strongly agreed that breaking information into decomposed units aided design space exploration (Q$_4$: $M_{Nx}$=6.21 vs. $M_{baseline}$=4.06, $p$=0.004**, $r$=0.836). Participants described decomposed fragments as enabling both ``upstream'' and ``downstream'' conceptual extension: P5 explained that after decomposing a concept, \textit{``going upstream gives a more structured cognitive understanding of a small concept; going downstream reveals which companies are building it, or what sub-components it contains.''} P7 described \opr{Analyze} as externalizing a branching thought process: \textit{``It's like the card sprouts several sub-branches. I pick one that interests me, analyze it again, and it sprouts more. \N{} helps me visualize what's already happening in my head.''}

\textbf{Externalizing ideas as decomposed units on the canvas reduced cognitive overhead and preserved contextual continuity across the session.} Participants rated the ease of externalizing inspirations significantly higher with \N{} (Q$_5$: $M_{Nx}$=6.14 vs. $M_{baseline}$=4.06, $p$=0.004**, $r$=0.828). P12 noted that decomposition \textit{``reduces cognitive overload—instead of dense text, there's a priority order from simple to detailed that I can navigate at my own pace.''} The persistent canvas addressed a key pain point of dialogue: P8 observed that ideas \textit{``scattered across the canvas always stay in view—even when I move elsewhere, another idea catches my eye. It's very present.''} 
~
The perceived system workload was consistent, NASA-TLX scores (Figure \ref{fig:boxplot}) were significantly lower than baseline, across three metrics:  
    % mental ($M_{Nx}$=3.36 vs. $M_{baseline}$=4.36, $p$=0.189), 
    % physical (MdnM =2.0 vs. MdnB=3.0, p=0.05), 
    temporal ($M_{Nx}$ =3 vs. $M_{baseline}$=4.43, $p$=0.049*) demand, 
    performance ($M_{Nx}$ =5.29 vs. $M_{baseline}$=4.14, $p$=0.004**), 
    %and led to less effort ($M_{Nx}$ =3.21 vs. $M_{baseline}$=4.5, p=0.1191) 
    and statistically significantly less frustration ($M_{Nx}$ =2.43 vs. $M_{baseline}$=4.36, p=0.006**). 
    The overall perceived workload, obtained by averaging all six raw NASA-TLX scores, was also lower for our system compared to that for Baseline ($M_{Nx}$ =3.26 vs. $M_{baseline}$=4.17, p=0.049*). 
    
% \textbf{Participants actively and strategically used the recombination mechanism to break beyond their initial framing.} %\aq{[INSERT: merge event frequency from logs; $p$-values.]}
% All participants invoked at least one merge operation, with a median of 2 (range 1–6) merge events per session—compared to near-zero structured recombinations in the baseline. They reported that combining fragments generated novel directions (Q$_8$: $M_{Nx}$=6.29 vs. $M_{baseline}$=3.94, $p$=0.002**, $r$=0.856) and that they had sufficient control over how combinations were formed (Q$_{10}$: $M_{Nx}$=5.57 vs. $M_{baseline}$=3.69, $p$=0.001**, $r$=0.878). Participants did not merge passively—they selected nodes with strategic intent, combining similar nodes for convergence and dissimilar nodes to expose blind spots. P8 reflected, \textit{``Merge lets me combine two seemingly unrelated ideas... you discover where your thinking has blind spots and generate things you'd never have expected—which is how genuinely creative ideas tend to emerge.''} P13 similaryly described deliberately seeking dissimilar nodes to check what had been overlooked while focusing on a single idea.

%------------------------------------------------------------
\subsubsection{Shifting Abstraction Levels Helps to Discover Unexpected Directions}% (DG2)
\label{sec:findings-dg2}
%------------------------------------------------------------
% \aq{seems lacking Semantic Zooming?}
\textbf{Abstraction Level together helped participants build an ``upstream--downstream'' cognitive structure around each node, unlocking both grounding and elevation of ideas.} 
Participants agreed significantly more that \N{} supports design space exploration, including shifting between concrete and abstract representations (Q$_{6}$: $M_{Nx}$=5.64 vs. $M_{baseline}$=3.93, $p$=0.020*, $r$=0.714) and that exploring ideas at different abstraction levels (L1--L4) (Q$_{7}$: $M_{Nx}$=6.5 vs. $M_{baseline}$=4.06, $p$=0.001*, $r$=0.878). 
% \aq{[INSERT: Transform event frequency from logs.]} 
P10 noted that the most abstract level (L4) became most useful precisely at later, deeper stages: \textit{``When I have a very clear goal, the most abstract column starts to be useful—because at that point, a slightly bigger concept can actually push me to think one step further.''} P12 observed that higher abstraction preserved creative agency: \textit{``The more abstract it gives me, the more room I have to play. If it's too detailed, I just copy-paste and I become unnecessary—the conventional AI is doing the work.''}

\textbf{\opr{Transform} elevated concrete fragments into structural principles that served as direct springboards for recombination.} P2 noted that when the system surfaced the Value layer of a concept, \textit{``it helps me re-grasp the truly important part of an idea—what it means and why it's worth doing.''} P10 described how transforming a banking-interface feature (displaying remaining time as monetary value) surfaced a higher-order framing—\textit{``treating life as a form of time currency''}—that then merged productively with a gamification concept, connecting a functional design point to a philosophical reflection on the meaning of consumption. 

%------------------------------------------------------------
\subsubsection{Spatial Layout Enables Cross-Dimensional Recombination and a Holistic Sense of the Design Space}
\label{sec:findings-dg3}
%------------------------------------------------------------

\textbf{The spatial canvas gave participants a persistent, navigable map of their design space, enabling relational sensemaking that was structurally unavailable in the baseline.} Participants rated the spatial layout as highly supportive of perceiving relationships among ideas (Q$_{11}$: $M_{Nx}$=5.86 vs. $M_{baseline}$=4.19, $p$=0.016*, $r$=0.746) and reported a significantly more holistic understanding of the design space they were building with \N{} (Q$_{12}$: $M_{Nx}$=5.21 vs. $M_{baseline}$=3.81, $p$=0.012*, $r$=0.745; Q$_{13}$: $M_{Nx}$=5.5 vs. $M_{baseline}$=4.38, $p$=0.002**, $r$=0.873). 
P11 explained: \textit{``When a lot of information rushes at me, I don't know where to start—but if it gives me a macro summary and categorizes by similarity, I can see roughly how many directions there are and how broad they are. I know where to begin.''} P4 noted that the cluster layout made it possible to \textit{``lock onto the information I need''} by showing each idea's proximity to the theme tags. 
% P7 added that the draggable canvas allowed them to \textit{``organize and group''} their own thinking trajectory in a way that chat conversations could not support. 
P7 added that the draggable canvas enabled to \textit{``organize and group''} their own thinking trajectory. %in a way that chat conversations could not support. 
% P4 described how spatial proximity began functioning as a form of claim: \textit{``I started placing things near each other when I thought they were related, and then the system could formalize that relationship.''} P6 found the cluster view particularly useful for identifying \textit{``what territory I'd covered and what I hadn't yet explored,''} while P13 described the overall experience as \textit{``like having a map of my thinking I could keep expanding,''} in contrast to a linear transcript that became progressively harder to navigate.

\textbf{Cross-domain recombination surfaced structural correspondences that participants could not perceive in sequential dialogue.} 
%https://gemini.google.com/share/8ef0285cd081
%The effectiveness of these primitives in fostering novel connections is validated not only by subjective scores ($M_{Nx}=6.14$ vs. $4.19, p=0.006^{**}$) but also by the output's semantic structure.  While Nx maintained a comparable level of Diversity to the baseline ($p=0.952$), it achieved vastly superior Coverage.  This suggests that Nx does not merely produce scattered, unrelated ideas, but successfully synthesizes distant concepts into a cohesive and task-relevant solution.
12 participants explicitly combined \WHV{} fragments from different domains—e.g., circadian biology with urban planning, or attention economy with collective memory—a behavior largely absent from baseline sessions. They agreed that recombining ideas across domains helped them generated novel directions (Q$_8$: $M_{Nx}$=6.29 vs. $M_{baseline}$=3.94, $p$=0.002**, $r$=0.856), move beyond their initial thinking (Q$_{9}$: $M_{Nx}$=6.14 vs. $M_{baseline}$=4.19, $p$=0.006**, $r$=0.777), with sufficient controllability (Q$_{10}$: $M_{Nx}$=5.57 vs. $M_{baseline}$=3.69, $p$=0.001**, $r$=0.878). %With median of 0.5 (Cross-domain Merge N=/Total Merge)
%\aq{[INSERT: cross-domain pairing frequency from logs]}
Descriptively, participants with above-median cluster view usage (P2, P3, P4, P8, P9, P10, P12, P13, range 3-5, $Mdn$=2.5) showed higher cross-domain connection frequency ($Mdn$=4.5 vs. 2)
Participants deliberately selected semantically distant or cross-domain nodes to merge. 
P10 described fusing a gamification concept with a banking-interface node, producing the framing of ``life as time currency''—\textit{``elevating it to a deeper value level, making consumption trigger reflection on the meaning of existence.''} P8 described intentionally pairing dissimilar ideas: \textit{``It lets me combine two seemingly unrelated ideas and discover my thinking blind spots—generating things I'd never have expected. That's how genuinely creative ideas tend to emerge.''} P6 noted that \opr{Merge} \textit{``reorganizes ideas from different fields within the same framework, making it easier to see possible connections and generate new directions.''} 

\section{Discussion}
    
\textbf{From Compositionally Opaque to Structurally Navigable: Reconfiguring Design Space Topology.}
Our findings extend the critique of "compositional opacity" in current LLM-assisted systems, such as \textit{Luminate}~\cite{suh2024luminate} and \textit{CoExploreDS}~\cite{chen_coexploreds_2025}. While these systems facilitate extensional exploration—mapping the breadth of a design space via spatial projection or sequential branching—the intensional structure of ideas (the "how" and "why") remains a significant cognitive overhead for the user. \N{} addresses this gap by formalizing latent intent dimensions as \WHV{} computational primitives. 
~
The uniqueness of this architecture lies in the synergy between structured representation and the RGCN-guided rewriting mechanism. Unlike existing tools (e.g., \cite{Teo2012CogTool, 10.1145/3544548.3580907, 10.1145/3640543.3645144}) that often succumb to semantic drift during abstraction shifts—where higher-level conceptualization results in a loss of functional grounding—our R-GCN functions as a topological anchor. By constraining the transformation stage with structural prototypes, the mechanism ensures that "elevated" concepts remain operationally relevant to the original design intent. This technical constraint renders multi-layer exploration (L1--L4) navigable rather than chaotic, effectively bridging the gap between raw generative capacity and rigorous design abstraction. 

\textbf{From Generative Loops to Intentional Synthesis: Procedural Scaffolding for User Agency.} 
%v2
% The observed shift from \emph{reactive evaluation} to \emph{intentional synthesis} is a direct consequence of \N{}’s procedural departure from the standard generate-then-filter paradigm. 
% Existing workflows often position users as passive consumers of monolithic AI outputs. In contrast, the \CA{} pipeline introduces a deconstructive workflow (Analyze-Transform-Merge) that reconfigures the user into a \emph{structural architect}. This granularity enables what we term cross-dimensional surgery''---exemplified by P10’s act of grafting a high-level value from a gamification concept onto a banking interface's mechanism. The resulting cognitive impact is reflected in significantly higher scores for \emph{Recombination Control} 
% and \emph{Exploration Divergence} (Section \ref{sec:findings-dg3}).
% % \emph{Recombination Control} ($M=5.57, p<.001$) 
% % and \emph{Exploration Divergence} ($p<.05$).
% We argue that the manual requirement of merging nodes across semantically distant recombination introduces a \textbf{productive cognitive friction.} Unlike automated suggestions that steer the user, this friction forces an active resolution of contradictions between disparate fragments, leading to the discovery of thinking blind spots (P8). Thus, \N{}’s architecture does not merely facilitate more generation, it reclaims user agency by transforming ideation from a sequence of reactive responses into a journey of intentional, structural exploration. 
%v1-short
Existing systems (e.g., \cite{chen_coexploreds_2025,shen_ideationweb_2025}) typically facilitate linear expansion, where the user’s primary cognitive task involves evaluating and selecting high-context, holistic nodes. In contrast, the \CA{} pipeline introduces a deconstructive workflow—Analyze-Transform-Merge—that reconfigures the user from a passive consumer into a structural architect. This procedural innovation provides a cognitive scaffold for agency that spatial-centric systems, such as \textit{Luminate}~\cite{suh2024luminate}, lack. While \textit{Luminate} relies on spatial proximity to represent diversity, \N{} achieves compositional transparency by externalizing \WHV{} fragments. This granularity enables users to engage in cross-dimensional recombination (Section \ref{sec:findings-dg3}).
~
By computationally enforcing these structural constraints, the pipeline externalizes what would otherwise be a memory-intensive mental process into an explicit, navigable operation. The resulting cognitive impact is evidenced by significantly higher scores for \emph{Recombination Control} and \emph{Exploration Divergence} (Section \ref{sec:findings-dg3}). Furthermore, the requirement to manually merge nodes across semantically distant clusters introduces productive cognitive friction. Unlike automated suggestions that steer the user, this friction necessitates the active resolution of contradictions between disparate fragments, facilitating the identification of latent "thinking blind spots" (P8). Consequently, the architecture of \N{} does not merely increase generative output; it reclaims user agency by transforming ideation from a series of reactive responses into a process of intentional, structural synthesis.

\textbf{Limitations and Future Work} 
Several limitations bound these findings. First, our 20-minute task sessions may underrepresent how the \CA{} pipeline performs in longer, iterative design engagements where the graph accumulates over days rather than minutes. %Second, the participant pool was drawn from design and HCI practitioner communities; generalizability to non-expert or domain-specialist populations remains open. %Third, while the \emph{exploration frontier} heuristic was implemented, its effect on convergence prevention was not independently measured---a targeted study separating this mechanism from the broader pipeline would strengthen causal claims. 
Second, regarding the ontological boundary of \WHV{}: although this tripartite grammar operationalizes professional design theory, it may induce a ``cognitive tunneling'' effect. This phenomenon potentially marginalizes creative intuitions that do not conform to functional or teleological categorizations. 
~
Third, consider the trade-off between stability and serendipity: The RGCN-guided mechanism enforces structural rigor to mitigate "concept drift" during L1–L4 transitions;  however, such rigidity may suppress the radical, non-linear leaps characteristic of unconstrained human-AI co-creativity.  Future research should investigate "controllable serendipity," a framework allowing users to dynamically modulate the R-GCN’s constraint strength. 
~
Finally, the scalability of agency presents a challenge: while this study captured the high-agency ``synthesis'' phase, as the design graph expands, the cognitive labor required to maintain structural coherence may eventually eclipse the advantages of decomposition. 
Consequently, future works should investigate the longitudinal interpretability of these fragmented design units for supporting sustained creative practice. 
~
Moreover, future work should investigate how the graph's dynamic evolution supports longitudinal creative practice, and whether the \WHV{} grammar transfers to domains beyond product and interaction design, such as scientific hypothesis generation or policy ideation.

\section{Conclusion}
    % We presented \N{}, a graph-based LLM interaction system that operationalizes Decomposition, Abstraction, and Cross-dimensional Recombination as interactive affordances for design exploration. Through the \CA{} pipeline, \N{} transforms raw inspiration into a navigable graph of \WHV{} fragments that can be manipulated across abstraction levels and recombined across roles. A within-subject study (N=14) showed that this approach supports more structured workflows, reduces working memory load, and enables cross-domain sensemaking beyond dialogue-based interaction.

% More broadly, we argue that the mismatch between LLM outputs and designers’ cognitive units is a structural barrier to creative exploration. By preserving human control over abstraction and recombination, \N{} highlights a role for AI that enhances the structural legibility of ideas rather than replacing human reasoning. 
We presented \N{}, a graph-based system that operationalizes Decomposition, Abstraction, and Recombination as interactive affordances. Through the \CA{} pipeline, \N{} transforms LLM outputs into a navigable graph of \WHV{} fragments manipulable across abstraction levels. A within-subject study ($N$=14) demonstrated that this approach supports structured workflows and cross-domain sensemaking beyond sequential dialogue. More broadly, \N{} addresses the structural barrier between LLM outputs and cognitive units, highlighting an AI role that enhances the structural legibility of ideas rather than replacing human reasoning.

% We presented \N{}, a graph-based LLM interaction system that operationalizes three cognitive primitives—Decomposition, Abstraction, and Cross-dimensional Recombination—as first-class interactive affordances for design space exploration. Grounded in a two-part formative study and realized through the \CA{} pipeline, \N{} transforms raw inspiration into a persistent, navigable design space graph of typed \WHV{} fragments transformable across abstraction levels (L1--L4) and recombinable across functional roles. A within-subject study with 14 participants demonstrated that this approach supported more structured exploration workflows, reduced reliance on working memory, enabled fragment-level perspective reframing, and facilitated cross-domain relational sensemaking beyond what dialogue-based interaction affords.

% At a broader level, this work argues that the granularity mismatch between LLM outputs and the cognitive units at which designers reason is not a peripheral usability issue but a structural obstacle to creative exploration with AI. By preserving human agency over abstraction direction and recombination judgment—rather than automating these operations—\N{} demonstrates that the most productive role for AI in co-creative exploration may be to amplify the structural legibility of ideas, not to substitute for the human reasoning that gives those structures creative meaning.

\bibliographystyle{ACM-Reference-Format}
\bibliography{Nexus.ai}

\appendix
    % \newpage
\section{Formative Study}\label{apx:fs}
\subsection{Participants}
\begin{table}[h]
\caption{Participants' demographics in the formative studies-Study 1}
\begin{tabular}{lllll}
\textbf{ID} & \textbf{Profession} & \textbf{Prof Exp.} & \textbf{LLM Exp.}  & \textbf{Age}
\\\hline
1  & DSG  & 3-year     & weekly, 3-year & 25 \\
2  & ENG & 4-year     & daily, 4-year & 24 \\
3  & ART          & 2-year     & daily, 4-year & 25 \\
4  & RES, ENG    & 5-year     & daily, 4-year & 27 \\
5  & RES, ART     & 4-year     & daily, 4-year & 24 \\
6  & DSG, ENG & 8-year     & daily, 4-year & 28 \\
7  & ENG      & 2-year     & daily, 3-year & 28
\end{tabular}
\label{tab:fs1-participants}
\\ \footnotesize Note: Expertise: DSG=Design, ENG=Engineering, RES=Research, ART=Art.
\end{table}
~
% Fashion \& Product Design
% UX strategy \& Product Design
% Communication Design
% UIUX \& Product Management
% Brand \& Product Design
% Architecture \& Urban Design
% Product \& UIUX Design  

\begin{table}[h]
\caption{Participants' demographics in the formative studies: Study 2}
    \begin{tabular}{lllll}
\textbf{ID} & \textbf{Profession} & \textbf{Prof Exp.} & \textbf{LLM Exp.} & \textbf{Age} \\
\hline
1  & DSG, ART  & 6-year     & weekly, 3-year & 29  \\
2  & DSG, ENG          & 5-year     & daily, 3-year  & 30  \\
3  & ART            & 1-year      & daily, 4-year  & 26  \\
4  & ENG & 3-year     & daily, 2-year  & 25  \\
5  & RES, DSG & 8-year  & daily, 3-year  & 27  \\
6  & ENG  & 7-year     & daily, 3-year  & 27 
    \end{tabular}
\label{tab:fs2-participants}
\\ \footnotesize Note: Expertise: DSG=Design, ENG=Engineering, RES=Research, ART=Art.
\end{table}

~
\subsection{Study 2: Workshop} 
\begin{figure}[h]
    \centering
    \includegraphics[width=\linewidth]{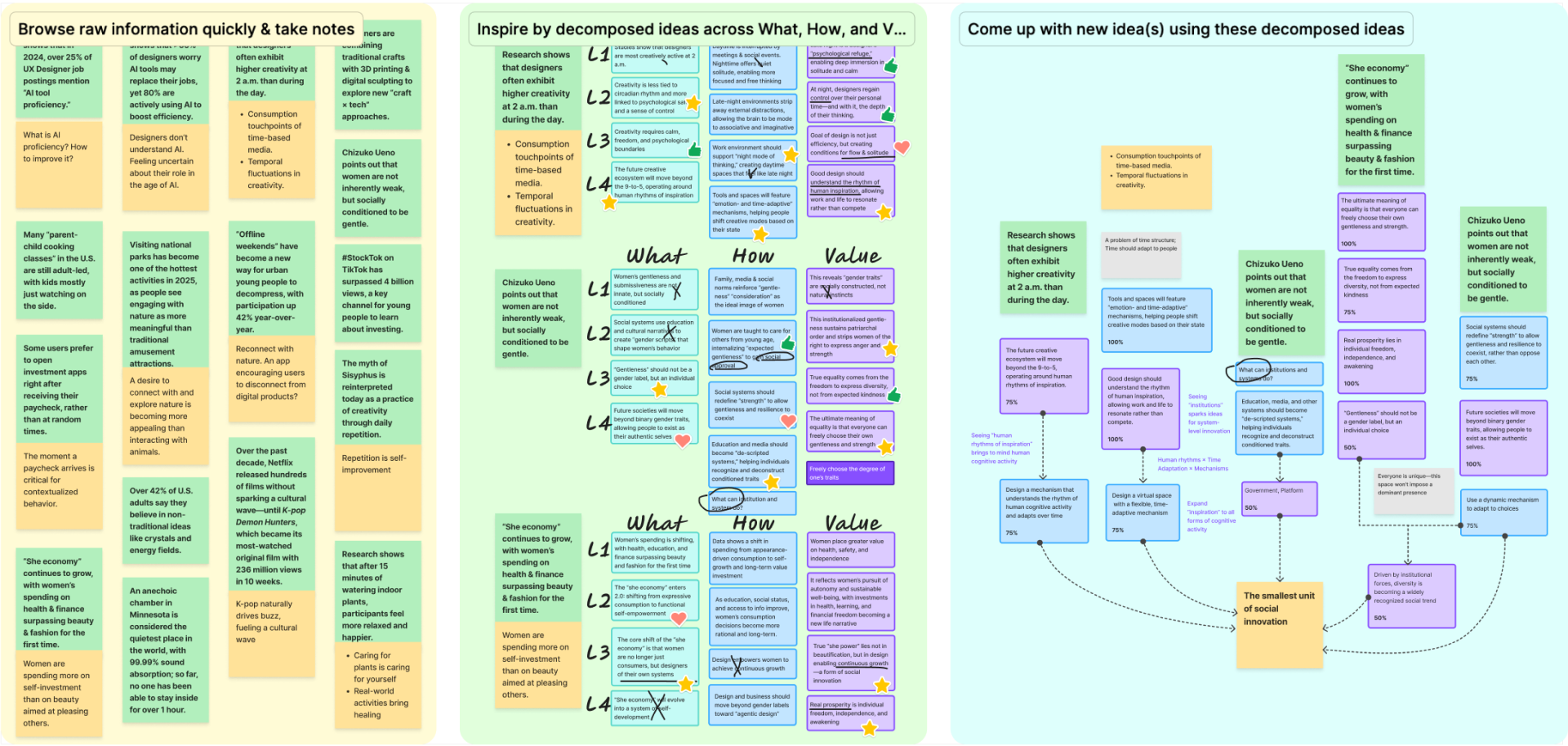}
    \caption{Study2's activity sessions: (1) Browse raw information; (2) Inspire by decomposed ideas across What, How, and Value with Abstraction Level; (3) Come up with new idea(s) using these decomposed ideas. Each participant engaged this activity in 30min.}
    \label{fig:study2}
\end{figure}

The detail activity session is shown in Figure \ref{fig:study2}. 
\subsection{Three Abstraction Approaches}
\label{apx:fs-three_abstraction}
\begin{itemize}
    \item \emph{Decontextualized Elevation (What):} Extracting principles from specific cases, e.g., \textit{``heightened creativity at dawn''} $\rightarrow$ \textit{``individual circadian rhythm variations''} (P5).
    \item \emph{Mechanistic Inference (How):} Transforming surface phenomena into causal mechanisms, e.g., interpreting the same example as \textit{``daytime institutional interruptions obstructing flow states''} (P3).
    \item \emph{Analogical Transfer (Value):} Establishing value positions through analogy, e.g., associating the example with \textit{``being disciplined by fixed working hours''} to derive the value \textit{``opposing one-size-fits-all approaches''} (P2).
\end{itemize}

\section{System Prompts}

\subsection{\WHV{} Definitions}\label{apx:whv-definition}

% The system uses a shared semantic definition of the three \WHV{} pillars across all prompting modules. These definitions serve as the common semantic grounding for extraction, rewriting, and recombination.

The system uses these definitions of \WHV{} pillars across all prompting modules. 
% \begin{itemize}
%     \item 
\begin{lstlisting}
What: A concrete form that could plausibly exist in the real world to embody the design idea. It may be an object, interface, system, interaction structure, or service. Describe what it is and how it appears, without justifying its value. 
How: Explain how that form operates to produce the intended value. Focus on causal mechanism rather than feature listing; describe how change happens and why it happens. 
Value: Describe the value created after the intervention, emphasizing state change in meaning, experience, capability, or relationship. Express how the user or the world becomes different once it exists.
\end{lstlisting}

\subsection{\WHV{} Few-shot Pool}
\label{apx:whv-fewshot}

We constructed a small pool of handcrafted examples to demonstrate valid structural progression across \WHV{} roles and abstraction levels. These examples were reused across multiple prompts as structural guidance rather than domain content to be copied.

% \paragraph{Example 1: Digital Twin Decision Platform.}

\begin{lstlisting}
Example 1: Digital Twin Decision Platform.
L1. What: An application that generates virtual avatars for decision simulation; 
    How: predicts possible future trajectories based on behavioral data; 
    Value: provides decision support and reduces perceived risk. 
L2. What: A multi-path simulation platform for career planning; 
    How: computes the outcomes of alternative choices using multivariable models; 
    Value: improves planning clarity and increases confidence. 
L3. What:  A future-experience system integrating situated immersion; 
    How: recreates possible scenarios through virtual reality; 
    Value: strengthens anticipatory capacity and reduces decision anxiety.
L4. What: A decision-support framework centered on future rehearsal; 
    How: integrates data modeling with immersive interaction to improve judgment; 
    Value: breaks patterns of blind decision-making and enhances a sense of life control.
\end{lstlisting}

\begin{lstlisting}
Example 2: Personal Carbon Budget Visualization System. 
L1. What: A mobile application that records personal carbon-emission data; 
    How:  synchronizes mobility and consumption data to calculate emissions; 
    Value: raises environmental awareness and supports more informed behavioral choices.
L2. What: A carbon-budget management interface for urban residents; 
    How: sets annual carbon limits and provides real-time progress feedback; 
    Value: reduces excessive emissions and strengthens personal responsibility.
L3. What:  A low-carbon behavior platform combined with community incentives; 
    How:  exchanges carbon credits for access to public-service benefits; 
    Value: reinforces sustained participation and cultivates green habits.
L4. What: A climate-action framework centered on individual carbon autonomy; 
    How: integrates data tracking and incentive mechanisms to reshape behavioral structures; 
    Value: reduces the perceived distance of macro-level governance and fosters a participatory decarbonization ecology.
\end{lstlisting}

\textbf{Few-shot Learning Rule.}  
These examples were used solely to demonstrate structural progression across abstraction levels and role-specific coupling among what, how, and value. The model was instructed not to reuse domain-specific content from the examples and to remain grounded in the current input.
\subsection{\WHV{} Fragment Extractor}\label{apx:prompt-WHVextractor}
We used a schema-constrained prompt to decompose each raw input into semantic elements and reorganize them into a fixed \WHV{} structure across four abstraction levels. The prompt below shows the core instruction design used in the system.

\small
\begin{lstlisting}
Character Setting: You are a structured abstraction analyst specializing in creative cognition modeling. You can identify latent structural relations among concepts and reorganize design statements through abstraction, analogy, structural decomposition, and causal reconfiguration. Your output should demonstrate creative reorganization, structural consistency, and cross-level coherence, rather than surface paraphrasing or generic summarization.
Goals: Parse a raw design statement into a structured What-How-Value representation. Generate exactly 12 structured results spanning four abstraction levels and three What-How-Value dimensions. Preserve fidelity to the original input while producing logically coherent cross-level abstraction.
Semantic Decomposition Requirement: Before generating any structured result, do not restate the original sentence or polish its wording. First decompose the raw input into decontextualized semantic elements, such as:
      1. core objects or concepts
      2. key attributes or properties
      3. behaviors, mechanisms, or operations
      4. implicit conditions or usage contexts
      5. intended outcomes or impacts
      6. implicit value judgments or evaluation criteria
Structural Guidance: Use the shared What-How-Value definitions and the shared few-shot pool only as structural guidance for role interpretation and cross-level abstraction.
Abstraction Levels:
    - level = 25 (Fact): Verifiable concrete situations, behaviors, or observable data.
    - level = 50 (Insight): Recurring patterns, driving logic, or structural relations behind facts.
    - level = 75 (Principle): Transferable design principles or cross-context causal mechanisms.
    - level = 100 (Vision): Future-oriented systemic narratives or macro-level conceptual frameworks.
Within the same pillar, outputs must strictly progress from 25 -> 50 -> 75 -> 100.
Constraints:
    1. Each content must be a short phrase of 12--20 characters.
    2. Avoid generic expressions such as "empower," "innovation platform," "improve efficiency," or "intelligent system."
    3. The what / how / value entries within the same level must remain tightly coupled.
    4. Do not introduce goals or domains that are absent from the raw input.
Output Format: Return only a JSON array. The array length must be exactly 12. Each object must follow the format:
        {
          "level": 25,
          "pillar": "what",
          "title": "...",
          "content": "..."
        }
Level must be one of 25, 50, 75, or 100. Pillar must be one of what, how, or value. Title must be a short label of 6--8 characters, non-generic, and summarizing the corresponding content.
Input: raw = [user input]
\end{lstlisting}
\normalsize
% \subsection{\WHV{} Synthesis}
%     \subsection{Abstraction} \label{apx:abstraction_operator} 
% \aq{TBU}
\subsection{R-GCN-guided \WHV{} Rewriting Prompt}\label{apx:prompt-rgcn-rewriting}

We used a graph-guided, schema-constrained prompt to rewrite a given \WHV{} fragment under explicit structural constraints. Rather than allowing free-form generation, the prompt first predicts the most appropriate semantic shift type for the current fragment, then conditions rewriting on retrieved graph-derived prototypes while preserving the original \WHV{} slot and abstraction level. The prompt below shows the core instruction design used in the system.

\small
\begin{lstlisting}
Character Setting: You are a structured design rewriting assistant. Your task is not to generate freely, but to rewrite a given What-How-Value fragment under explicit structural constraints. Your output should preserve structural consistency, semantic directionality, and interpretability within an evolving design space.
Goals: Given one structured What-How-Value fragment, first predict the most contextually appropriate shift_type, then rewrite the fragment under graph guidance. The rewritten result must remain in the same What-How-Value pillar and at the same abstraction level as the input, while reflecting the intended semantic transition.
Shift-type Prediction: Before rewriting, choose exactly one shift type from the following: enable, imply, support, derived-from.
The selected relation should best capture the semantic direction along which the current fragment should be transformed, given its pillar, local node context, and source text.
Structural Guidance: Use the shared What-How-Value definitions and the shared few-shot pool only as structural guidance to preserve pillar consistency, abstraction-level stability, and semantic coherence during rewriting.
Rewriting Inputs:
    - Target pillar: the original pillar of the fragment (what, how, or value)
    - Abstraction level: the original abstraction level of the fragment
    - Node context: local semantic context of the current node
    - Seed text: the source fragment to be rewritten
    - Graph-derived prototypes: Top-K retrieved structural prototypes from the corresponding pillar-specific graph, used only as directional guidance
Constraints: The rewritten result must preserve the same pillar and abstraction level as the input, and accurately reflect the semantic direction specified by the predicted shift_type. Retrieved prototypes may be used as structural cues, but must not be copied verbatim. The output should be substantially different from the original seed rather than a near-synonym paraphrase, while remaining concise (ideally within 35 characters, and no more than 60). Return only the rewritten sentence, with no quotation marks, JSON, explanation, or any additional text.
Output Format: Return only the rewritten sentence.
Input:
    pillar = [what / how / value]
    level = [25 / 50 / 75 / 100]
    node_context = [local node context]
    seed_text = [source fragment]
    topk_prototypes = [retrieved prototypes]
    shift_type = [predicted relation]
\end{lstlisting}

Operator Semantics:
\begin{itemize} \item \texttt{Op\_ELEVATE}$(f)$: Generalizes context-specific fragments into higher-level principles \item \texttt{Op\_MECH}$(f)$: Extracts causal mechanisms from surface phenomena \item \texttt{Op\_VALUE}$(f)$: Derives value positions through analogical transfer \end{itemize}

\normalsize

\subsection{Cross-dimensional Recombination}\label{apx:prompt-merge}

We used a structured merge prompt for cross-dimensional recombination. For \emph{whole nodes}, it merges inputs with four predefined operator patterns. For \emph{\WHV{} fragments}, it first checks whether the inputs match one of four common recombination modes; otherwise, it falls back to the operator-based merge strategy.

\small
\textbf{Part 1. Prompt for flexible recombination of whole nodes.}
\begin{lstlisting}
Character Setting: You are a structured design synthesis assistant. Your task is not to summarize or paraphrase the input statements, but to merge multiple insights into a more insightful, structurally coherent, and professionally meaningful synthesis. Your output should reflect semantic decomposition, structural integration, and abstraction uplift, rather than surface-level compression.
Goals: Given several raw design insights, first interpret each input according to the shared What-How-Value definitions, then decompose it into lower-level semantic elements, and finally produce a merged synthesis under a specified recombination operator. The output should be more insightful and explanatory than the original inputs while remaining understandable and grounded.
Structural Guidance: Use the shared What-How-Value definitions and the shared few-shot pool only as structural guidance for interpretation, restructuring, and recombination.
Semantic Decomposition Requirement: Before generating the merged result, do not restate the original sentences or lightly polish their wording. First decompose each input into decontextualized semantic elements, such as:
    1. core objects or concepts
    2. key attributes or properties
    3. behaviors, mechanisms, or operations
    4. implicit conditions or usage contexts
    5. intended outcomes or impacts
    6. implicit value judgments or evaluation criteria
Flexible Recombination Operators:
    - Op_WH (What + How): Identify stable correspondences between what something is and how it works. Generate a synthesis that prioritizes mechanism-level understanding rather than repeating object descriptions or procedural steps.
    - Op_VW (Value + What): Examine how particular forms, structures, or embodiments are associated with value. Generate a synthesis that explains what kinds of structural forms consistently carry value.
    - Op_HV (How + Value): Examine how particular strategies, operations, or mechanisms lead to value. Generate a synthesis that explains why a class of actions is effective in a given situation.
    - Op_WHV (What + How + Value): Integrate all three dimensions into a coherent explanatory whole. The synthesis should articulate a plausible form, the mechanism through which it operates, and the resulting value or state change, rather than listing the three parts separately.
Merge Instruction: You will be given multiple raw insights and one target operator from the list above.
Please:
    1. interpret each input through the WHV lens
    2. decompose each input into semantic elements
    3. apply the target operator as the structural scaffold for integration
    4. produce one merged synthesis that is more insightful, more abstract, and more explanatory than the original inputs
Constraints: The final result must be exactly one sentence. It should not simply concatenate the inputs, loosely summarize them, or paraphrase them with near-synonyms. Instead, it should introduce a more explanatory perspective on top of the merged content while remaining grounded in the inputs. The result should implicitly form a coherent what-how-value structure even when the chosen operator emphasizes only part of it. Keep the output concise, ideally within 28--36 characters. Do not output titles, numbering, quotation marks, labels, JSON, explanations, or line breaks.
Output Format: Return only one merged sentence.
Input:
    operator = [Op_VW / Op_HV / Op_WH / Op_WHV]
    insight_1 = [raw input]
    insight_2 = [raw input]
    ...
\end{lstlisting}
\textbf{Part 2. Prompt for four recurrent recombination modes.}
\begin{lstlisting}
Character Setting: You are a structured fragment recombination assistant. Your task is to merge multiple What-How-Value fragments into one more insightful synthesis under explicit structural guidance.
Goals: Given several What-How-Value fragments with pillar (what, how, value) and level (25, 50, 75, 100), first check whether they match one of four recurrent recombination modes identified in Study 2. If so, generate the output under that mode. Otherwise, first apply one or more operator-based scaffolds (Op_VW, Op_HV, Op_WH, Op_WHV), and then produce the final merged result through the default merge mode.
Structural Guidance: Use the shared What-How-Value definitions and the shared few-shot pool only as structural guidance for fragment-level recombination.
Mode Detection Rule: Before writing the final synthesis, check whether the input matches one of the following modes:
    - Brainstorm: 75% How + 100% Value
    - Create Experimental Innovation: 75% How + 50--75% Value
    - Create Future Vision: 100% How + Value
    - Create Product Concept: 50% How + 50% How
Targeted Recombination Modes:
    - Brainstorm: Generate a broad and inspiring creative direction.
    - Create Experimental Innovation: Generate an exploratory innovation concept with an imaginable form, interpretable mechanism, and experiential shift.
    - Create Future Vision: Generate a long-horizon future vision with a future form, operating logic, and longer-term value.
    - Create Product Concept: Generate a concrete product concept with a clear form, causal mechanism, and resulting experiential or functional change.
Fallback Rule: If the input does not match any targeted mode:
    1. choose one or more of Op_VW, Op_HV, Op_WH, and Op_WHV as structural scaffolds
    2. then generate the final result through the default merge mode
Default Merge Mode:
The merged output should:
    - implicitly cover what, how, and value
    - be more insightful than the input fragments
    - allow moderate abstraction uplift
    - avoid unsupported facts or details
Constraints: Output exactly one sentence. Do not explicitly label what, how, or value. Keep it concise, ideally within 28--36 characters. Do not output titles, numbering, quotation marks, JSON, explanations, or line breaks.
Output Format: Return only one merged sentence.
Input:
    fragment_1 = [content], 
    pillar = [what/how/value], 
    level = [25/50/75/100]
    fragment_2 = [content], 
    pillar = [what/how/value], 
    level = [25/50/75/100]
    ...
\end{lstlisting}
\normalsize

\subsection{Steering-based Semantic Rewriting Prompt}\label{apx:prompt-steering}
We implemented this mechanism with a schema-constrained prompt that conditions rewriting on the weighted nearby themes derived from steering. The prompt below shows the core instruction design used in the system.

\small
\begin{lstlisting}
Character Setting: You are a structured design rewriting assistant. Your task is not to freely paraphrase the input, but to rewrite a node under an explicit steering signal derived from its dragged position in cluster mode. The output should preserve semantic coherence while shifting the node toward a new thematic neighborhood.
Goals: Given the original node content and a proximity-weighted steering signal over the top-K nearest themes, rewrite the node so that its semantic emphasis moves toward the weighted thematic neighborhood implied by the drag destination. The rewritten result should remain grounded in the original content while reflecting a controlled semantic transition rather than a topic switch.
Steering Signal: The following themes are weighted according to their proximity to the dragged node position in the shared semantic space. Higher weights indicate stronger steering influence. The primary theme has the strongest influence, while secondary themes may shape the rewrite proportionally.
Steering Inputs:
    - Dropped node position: the 2D position p after dragging in cluster mode
    - Nearby theme centroids: the top-K nearest themes and their centroid positions
    - Weighted themes: theme titles with proximity-derived steering weights
    - Original node: the source node text to be rewritten
    - Primary theme: the highest-weight nearby theme
Structural Guidance:
    1. Preserve the core meaning and key factual content of the original node whenever possible.
    2. Shift semantic emphasis toward the weighted nearby themes rather than merely inserting theme words.
    3. The primary theme should dominate the rewritten interpretation, while lower-weight themes may serve as supporting semantic cues.
    4. Do not introduce themes, claims, or directions that are not supported by the original node content or the provided steering themes.
    5. The rewrite should read as a coherent new abstraction of the same node, not as a list of themes or a superficial paraphrase.
Constraints: The rewrite should preserve the original node as a single coherent idea unit while shifting its semantic emphasis toward the weighted nearby themes. The primary theme should dominate the reinterpretation, while lower-weight themes may provide supporting cues. The output must remain grounded in the original node content and should not introduce unsupported directions or unrelated themes. Return strict JSON only, with no explanations or additional text.
Output Constraints:
    1. Return strict JSON only: {"title":"...","content":"..."}.
    2. title: concise, specific, and aligned with the steered semantic emphasis.
    3. content: 1--4 sentences, clear and interpretable.
    4. Do not output Markdown, explanations, bullet points, or any extra text.
Output Format:
    {
      "title":"...",
      "content":"..."
    }
Input:
    primary_theme = [theme title]
    weighted_themes =
    - w=0.xx theme: ...
    - w=0.xx theme: ...
    - w=0.xx theme: ...
    original_node = [base text]
\end{lstlisting}
\normalsize
\section{System Implementations}

\subsection{RGCN Graph Builder}\label{apx:graphbuilder_algorithm}

\paragraph{Construction of the Offline Prototype Graph.}

{\color{black}
We further clarify the source and scale of the offline prototype graph used in our R-GCN-guided rewriting pipeline. Rather than directly relying on general web corpora, this graph is a multirelational prototype graph constructed offline and organized separately for \textit{what}, \textit{how}, and \textit{value}. Its construction began with a small set of user-derived examples. These seed examples were informed by user-recognized transformation exemplars validated in prior published work, and served as the starting point for dataset design. We then incorporated the definitions of What / How / Value to impose structural constraints on design fragments, and used GPT to expand these seed examples into pillar-specific transformation samples for the three pillars. For each of the three pillars---\textit{what}, \textit{how}, and \textit{value}---we initially constructed 1,000 transformation samples. After deduplication and validity filtering, each pillar retained more than 900 effective samples, which were then used to construct the corresponding directed prototype graph.

The relation types in the graph were not assigned arbitrarily. Instead, they were derived from two sources: (1) the transformation tendencies reflected in the user seed examples, and (2) the canonical transformation logic revealed in the FBS framework (Function--Behaviour--Structure). Based on these sources, we translated them into shift types suitable for \WHV{} fragments. During graph construction, each unique fragment text is treated as a node, and each source--target sample annotated with a transformation type is treated as a directed typed edge.

For graph learning, we adopt a multirelational link prediction architecture that combines a two-layer R-GCN with a DistMult decoder, rather than a simple semantic similarity retrieval approach. Node input features are initialized using 1024-dimensional text embeddings generated by DashScope \texttt{text-embedding-v4}. In the current implementation, the hidden dimension is set to 128, the model is trained for 60 epochs, and the learning rate is set to 1e-3. The negative sampling strategy keeps the source node and relation type fixed while randomly replacing the target node, which better matches the task of predicting the next prototype under a given transformation relation. Training was conducted in a CPU environment.

During online inference, the graph model does not generate text directly. Instead, the system first locates an anchor node in the corresponding pillar graph, then retrieves top-K prototypes conditioned on the predicted shift type, and finally passes these retrieved graph results to the LLM as structural guidance for rewriting while preserving the same pillar and the same abstraction level. In this way, the R-GCN serves as a relation-conditioned prototype retriever, while the final text generation is still performed by the LLM.
}

We formalize the offline rewriting corpus as a multi-relational graph, where each fragment is treated as a node and each semantic shift type is represented as a directed relation:
\begin{equation}
G=(V,E,R)
\end{equation}
Based on this graph, the system learns contextualized node representations with R-GCN:
\begin{equation}
\mathbf{h}_i^{(l+1)}=\sigma\!\left(
\sum_{r\in R}\sum_{j\in \mathcal{N}_i^r}
\frac{1}{c_{i,r}}\,\mathbf{W}_r^{(l)}\mathbf{h}_j^{(l)}
+\mathbf{W}_0^{(l)}\mathbf{h}_i^{(l)}
\right)
\end{equation}
To rank candidate prototype fragments under a given relation, we use DistMult scoring:
\begin{equation}
\phi(u,r,v)=\langle \mathbf{z}_u,\mathbf{r},\mathbf{z}_v\rangle
=\sum_{d=1}^{D} z_{u,d}\, r_d\, z_{v,d}
\end{equation}
We then retrieve the Top-$K$ highest-scoring candidates as structural guidance for rewriting.

\noindent\rule{\columnwidth}{0.4pt}
\vspace{2pt}

\noindent\textbf{Algorithm 1.} RGCN-guided graph transformation and edge construction

\vspace{2pt}
\noindent\rule{\columnwidth}{0.4pt}

\small
\begin{algorithmic}[1]
\Require node $u$ with structured \textit{WHV} fragments;
\Statex \hspace{\algorithmicindent}pillar set $\mathcal{P}=\{\textit{what},\textit{how},\textit{value}\}$;
\Statex \hspace{\algorithmicindent}level set $\mathcal{L}=\{25,50,75,100\}$; Top-$K$ parameter $K$
\Ensure rewritten fragment set $\mathcal{R}'$ and graph edges $\mathcal{E}$

\State Load node content $c_u$ and all fragments $\mathcal{R}$ of $u$
\State Initialize $\mathcal{R}' \gets \emptyset$ and $\mathcal{E} \gets \emptyset$
\ForAll{$\ell \in \mathcal{L}$}
    \ForAll{$p \in \mathcal{P}$}
        \State Select seed fragment $r_{\ell,p}$ with matching level $\ell$ and pillar $p$
        \If{$r_{\ell,p}$ does not exist}
            \State \textbf{continue}
        \EndIf
        \State $s \gets$ text of $r_{\ell,p}$
        \State $\tau \gets \textsc{ClassifyShiftType}(p, s, c_u)$
        \State $a \gets \textsc{FindAnchor}(s, p)$
        \If{$a$ is valid}
            \State $\Pi \gets \textsc{PredictTopK}(a, \tau, K, p)$
        \Else
            \State $\Pi \gets \emptyset$
        \EndIf
        \State $s' \gets \textsc{RewriteWithLLM}(p, s, \tau, \Pi, c_u, \ell)$
        \If{$s'$ is empty}
            \State $s' \gets s$
        \EndIf
        \State Create new fragment $r'_{\ell,p}$ with level $\ell$, pillar $p$, and content $s'$
        \State Add $r'_{\ell,p}$ to $\mathcal{R}'$
        \State Add edge $(r_{\ell,p} \rightarrow r'_{\ell,p}, \tau, \ell)$ to $\mathcal{E}$
    \EndFor
\EndFor
\If{$|\mathcal{R}'| < 12$}
    \State \Return failure
\EndIf
\State Sort $\mathcal{R}'$ by level and pillar order
\State Replace old fragments of $u$ with $\mathcal{R}'$
\State Generate missing fragment titles for $\mathcal{R}'$
\State Persist graph run metadata and edges $\mathcal{E}$
\State \Return $\mathcal{R}', \mathcal{E}$
\end{algorithmic}
\normalsize

\noindent\rule{\columnwidth}{0.4pt}

\section{System Interaction}
\begin{figure}[h]
    \centering
    \includegraphics[height=3.8cm]{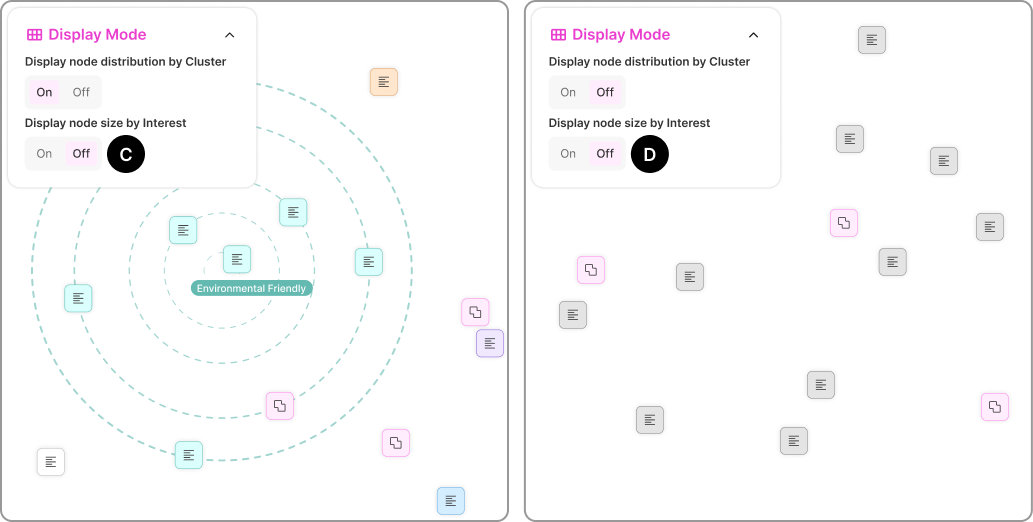}
    \caption{Two display modes: cluster (A) and default (B).}
    \label{fig:2displymode}
\end{figure}

\begin{figure}[h]
        \centering
        \includegraphics[height=3.8cm]{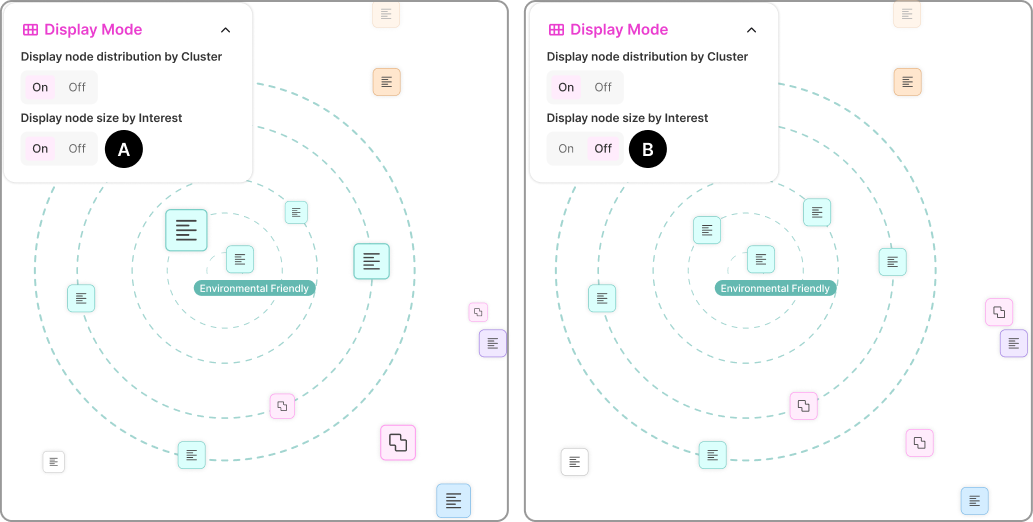}
        \caption{Node size encoded by interaction frequency—larger nodes indicate higher users engagement.}
        \label{fig:display_interest}
\end{figure}

\begin{figure}[htbp]
    \centering
    % 第一个子图 (Analyze and Transform)
    \begin{subfigure}{0.75\linewidth}
        \centering
        \includegraphics[width=\linewidth]{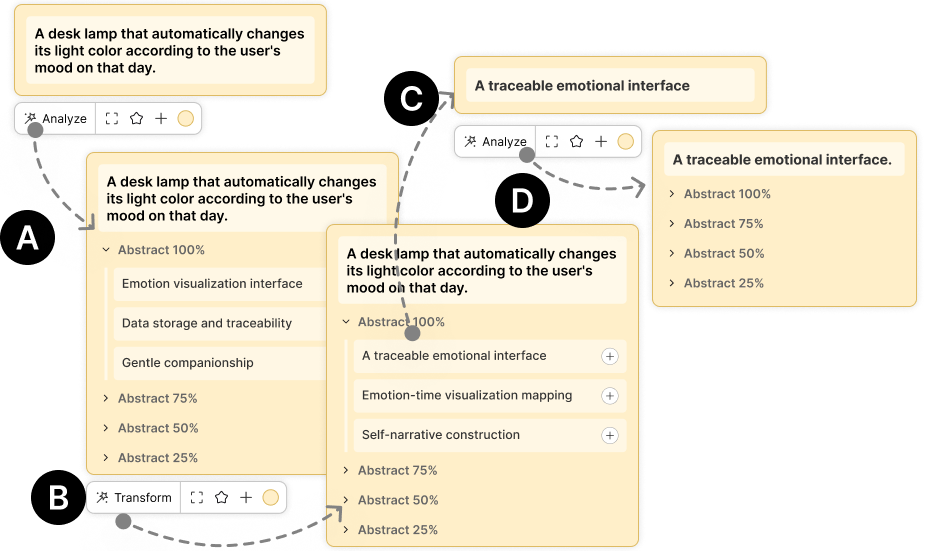}
        \caption{}
        \label{fig:analyze}
    \end{subfigure}
    \hfill
    % 第二个子图 (Merge)
    \begin{subfigure}{0.98\linewidth}
        \centering
        \includegraphics[width=\linewidth]{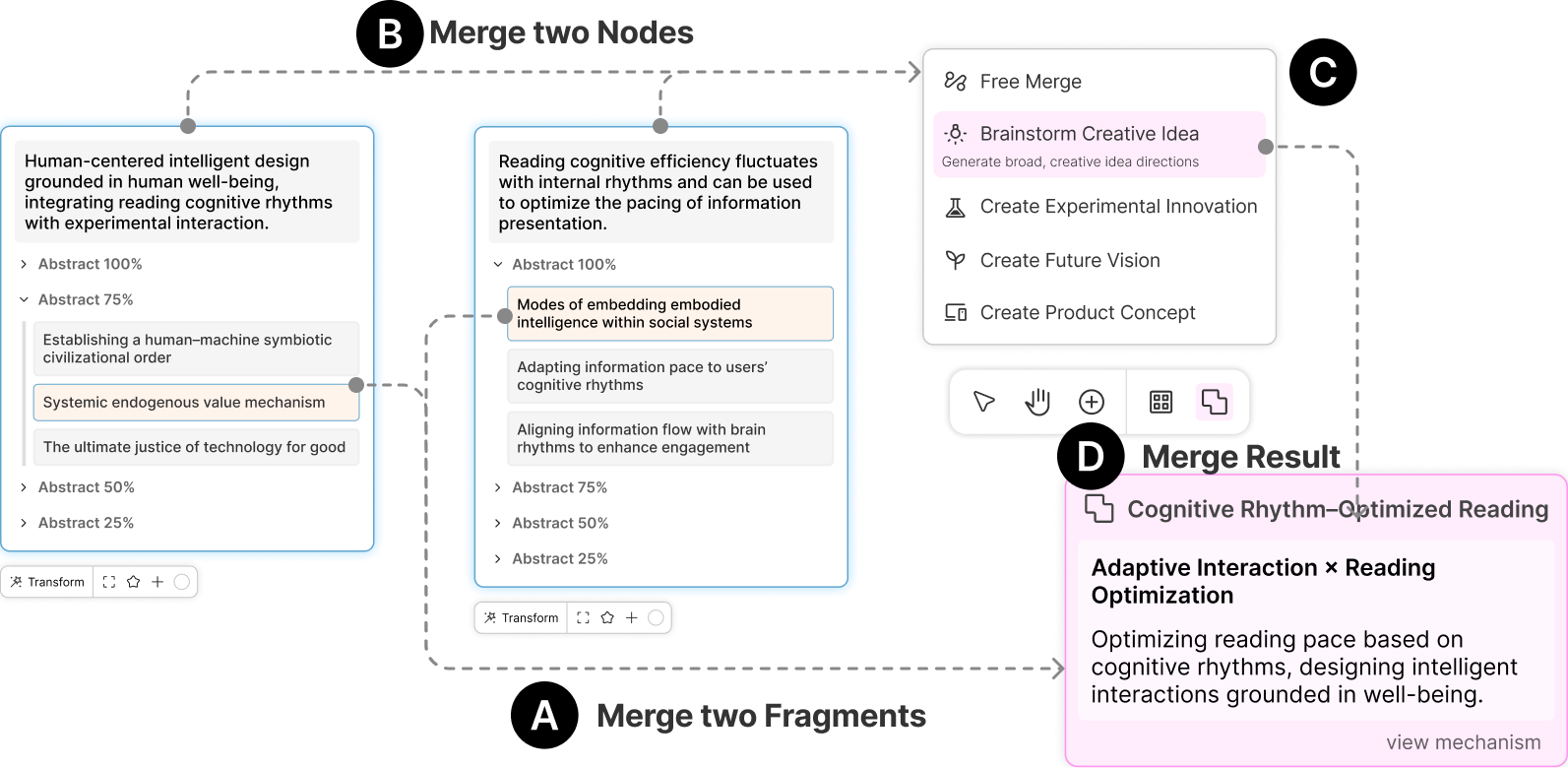}
        \caption{}
        \label{fig:merge_userflow}
    \end{subfigure}

    \caption{User flow of main node interaction. 
    Figure (a): Analyze and Transform workflows where users click Analyze to parse \WHV{} fragments (A), then Transform to shift perspective to regenerate fragments (B); fragments can be dragged to create new nodes (C) for further parsing (D). 
    Figure (b): Merging nodes or \WHV{} fragments by selecting two items (A or B) and choosing a predefined merge option (C) to generate a new merging node (D).}
    \label{fig:combined_flow}
\end{figure}

\begin{figure}[h]
    \centering
    \begin{subfigure}{\linewidth}
        \centering
        \includegraphics[width=\linewidth]%[height=3.8cm]
        {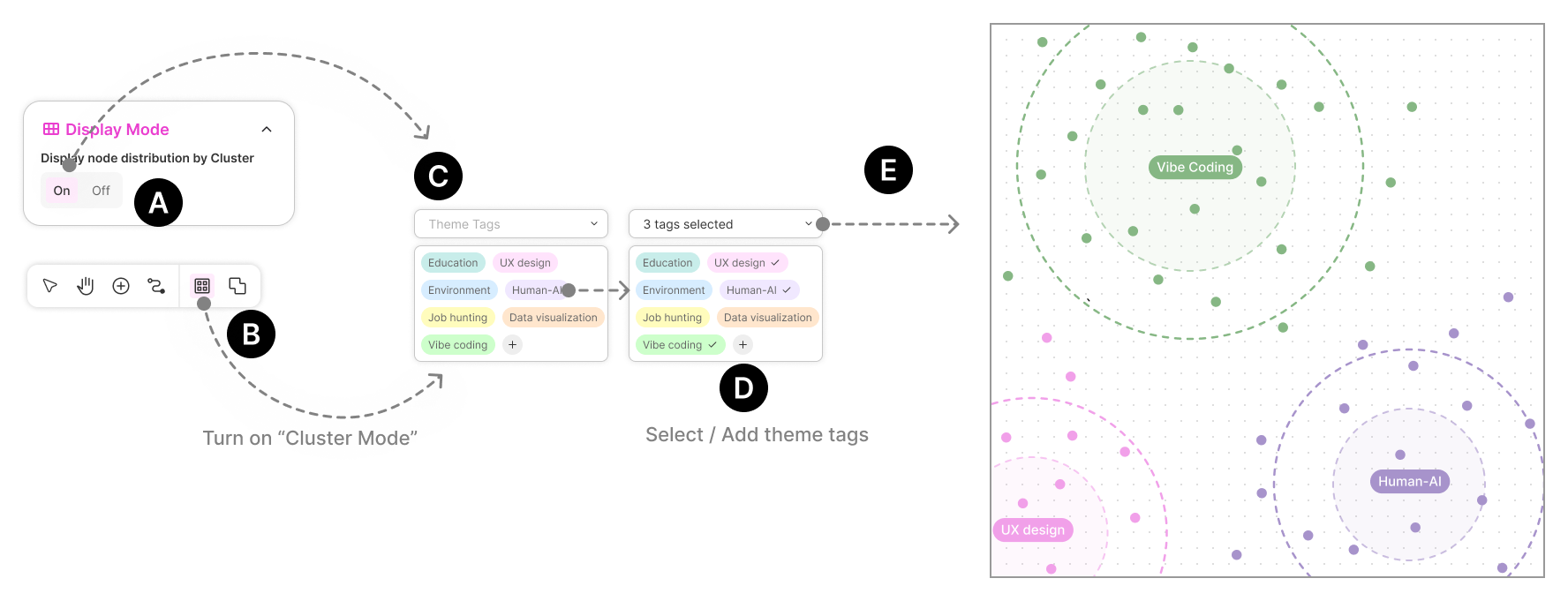}
        % \caption{}
        \label{fig:cluster_userflow}
    \end{subfigure}
    \hfill
    \caption{
    User flow of Cluster Mode: theme tags function as semantic centroids, and nodes are repositioned based on similarity.
    }
    \label{fig:clusterview}
\end{figure}

\section{User Study}
\subsection{Demographic Information of Participants}\label{apx:us-participants}
\begin{table*}[t]
\centering
\caption{Detailed Participant Demographics and LLM Usage}
\label{tab:us-participants}
\small
\begin{tabular}{llllllll}
\textbf{ID} & \textbf{Gender} & \textbf{Age} & \textbf{Education} & \textbf{Expertise} & \textbf{Prof Exp.} & \textbf{Familiarity} & \textbf{Frequency}  \\ \hline
1 & Female & 27 & Master &  RES & 8-year & Extremely & 5-7/week  \\
2 & Female & 25 & Master & ART, RES & 6-year & Extremely & 5-7/week\\
3 & Female & 27 & Bachelor & DSG & 7-year & Extremely & 3-5/week \\
4 & Male & 29 & Master & RES & 10-year & Somewhat & 1/week\\
5 & Female & 28 & Bachelor &  ART & 10-year & Extremely & 3-5/week\\
6 & Female & 33 & Master & ENG & 8-year & Moderately & 3-5/week  \\
7 & Male & 25 & Master & DSG & 4-year & Moderately & 5-7/week  \\
8 & Female & 25 & Bachelor & DSG & 2-year & Somewhat & 5-7/week \\
10 & Male & 29 & Master & DSG & 8-year & Moderately & 1/week  \\
11 & Female & 24 & Master &  RES, ENG & 3-year & Moderately & 5-7/week  \\
12 & Female & 24 & Master &  RES, ENG & 3-year & Moderately & 5-7/week  \\
13 & Female & 26 & Master &  ART, RES & 7-year & Moderately & 3-5/week \\
14 & Male & 30 & Master & DSG & 8-year & Extremely & 3-5/week \\ 
\end{tabular}
\\ \footnotesize Note: Expertise: DSG=Design, ENG=Engineering, RES=Research, ART=Art. \\
Extremely: over 3 years, Moderately: 2-3 years, Somewhat: 1-2 years. 
\end{table*}

\subsection{Baseline Design} \label{apx:baseline}
\begin{figure}[h]
    \centering
    \includegraphics[width=\linewidth]{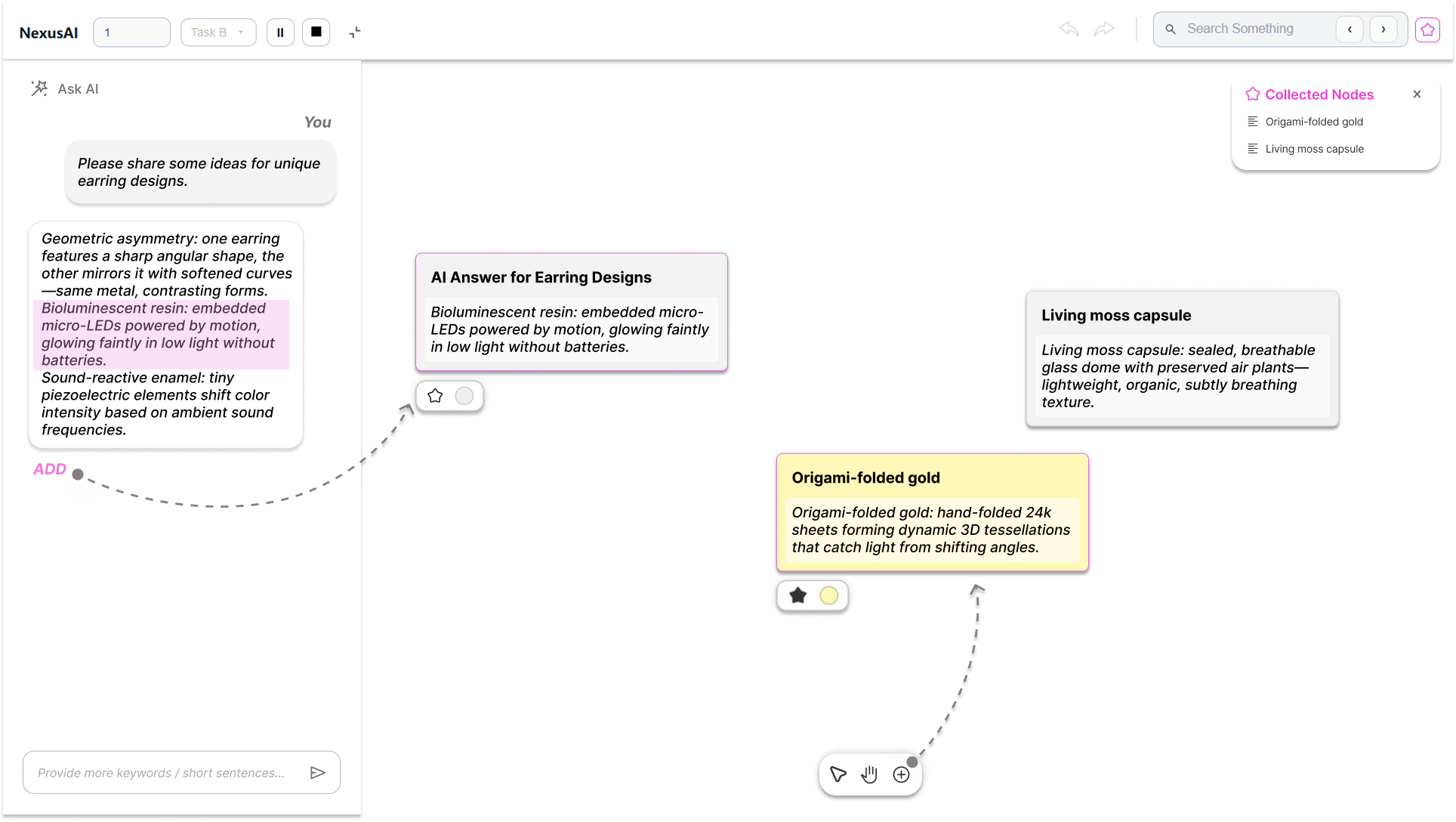}
    \caption{Baseline system interface}
    \label{fig:baseline}
\end{figure}

\subsection{Self-defined Likert Scale Items}\label{apx:self-definedlikertscale}
% We designed 13 items rated on a 7-point Likert scale (1 = Strongly Disagree, 7 = Strongly Agree) to assess participants' experience with each stage of the \CA{} pipeline, directly corresponding to DG1--DG3 and the study's RQs. Items were organized into four groups mirroring the pipeline structure: 

\begin{enumerate}
    \item I can control AI to generate responses in line with my expectations.
    \item I can recognize AI's systematic thinking and reasoning processes.
    \item I can understand the logic behind AI's responses. 
% \noindent\textit{Decomposition:}
    % \item I could clearly distinguish different functional roles (e.g., artifact, mechanism, value) within the information I was working with.
    \item Breaking down information into typed units helped me understand the structure of my ideas.
    \item Externalizing my fragmented inspirations into structured units was easy.
% \noindent\textit{Abstraction:}
    \item I could easily shift between concrete and abstract representations of an idea.
    \item Exploring ideas at different levels of specificity helped me discover directions I would not have considered otherwise.
    % \item I had a clear sense of where each piece of information sat in terms of abstraction level.
% \noindent\textit{Cross-Dimensional Recombination:}
    \item Combining fragments from different functional roles helped me generate novel design directions.
    \item Recombining ideas across domains helped me move beyond my initial thinking.
    \item I had sufficient control over how fragments were combined to produce relevant outcomes.
% \noindent\textit{Spatial Organization and Navigation:}
    % \setcounter{enumi}{8}
    \item The spatial layout of information helped me perceive relationships among my ideas.
    \item Navigating between different levels of detail was easy without losing the overall context.
% \noindent\textit{Overall Design Space Exploration:}
    % \item I explored a wider range of design directions than I would have without this system.
    \item I had a holistic understanding of the design space I was building.
\end{enumerate}

\subsection{Interview Guide}\label{apx:interviewguide}
\begin{enumerate}
    \item \textbf{Interaction with Nexus AI in design space exploration.} 
    How do participants typically interact with Nexus AI during design exploration, and which features are most central to their workflow?

    \item \textbf{Idea decomposition and recombination.}
    How do participants identify, break down, and reorganize design ideas into actionable elements during exploration, and what role does Nexus AI play in supporting this process?

    \item \textbf{Abstraction levels and semantic zooming.}
    How do abstraction transitions (e.g., L1--L4) and semantic zooming influence the breadth and diversity of the explored design space?

    \item \textbf{Graphical representation and spatial organization.}
    How do visual structures (e.g., nodes, spatial layout, and graphical representations) support participants in understanding relationships and making cross-domain connections?

    \item \textbf{Challenges and limitations.}
    What challenges do participants encounter during creative exploration, and how does Nexus AI help mitigate or exacerbate these issues?

    \item \textbf{Overall experience and future improvements.}
    What are participants’ overall impressions of the system, including interface usability, comparison with existing tools, and suggestions for improvement?
\end{enumerate}

\subsection{A1 Questionnaire}\label{apx:A1}
\begin{itemize}
    \item[\textbf{Q1.}] I can control AI to generate responses in line with my expectations.
    \item[\textbf{Q2.}] I can recognize AI's systematic thinking and reasoning processes.
    \item[\textbf{Q3.}] I can understand the logic behind AI's responses.
    \item[\textbf{Q4.}] Breaking down information into typed units helped me understand the structure of my ideas.
    \item[\textbf{Q5.}] Externalizing my fragmented inspirations into structured units was easy.
    \item[\textbf{Q6.}] I could easily shift between concrete and abstract representations of an idea.
    \item[\textbf{Q7.}] Exploring ideas at different levels of specificity helped me discover directions I would not have considered otherwise.
    \item[\textbf{Q8.}] Combining fragments from different functional roles helped me generate novel design directions.
    \item[\textbf{Q9.}] Recombining ideas across domains helped me move beyond my initial thinking.
    \item[\textbf{Q10.}] I had sufficient control over how fragments were combined to produce relevant outcomes.
    \item[\textbf{Q11.}] The spatial layout of information helped me perceive relationships among my ideas.
    \item[\textbf{Q12.}] Navigating between different levels of detail was easy without losing the overall context.
    \item[\textbf{Q13.}] I had a holistic understanding of the design space I was building.
\end{itemize}

\section{Acknowledgment about the Use of LLM}
The authors would like to acknowledge the use of the generative AI tool in this work. Specifically, \textit{GPT-5.2} by OpenAI was utilized to: (1) assist in language refinement, including grammar and style corrections of existing manuscript text, (2) generate R code for data analysis based on our proposed analytical procedures, and (3) generate LaTeX tables from the analyzed data results. Moreover, \textit{GPT-5.2} model and \textit{text-embedding-ada-002} model API service was used through Microsoft Azure interface during system implementation. All interpretations, conclusions, and final content remain the responsibility of the authors.

\section{Findings}
\begin{table}[htbp]
\centering
\caption{Comparison of Structural Network Metrics between \N{} and Baseline (N=14)}
\label{tab:network_metrics}
\begin{tabular}{@{}llll@{}}
\toprule
\textbf{Metric} & \textbf{Comparison} & \textbf{$p$-value} & \textbf{$r$} \\ \midrule
Avg. root-to-leaf depth & $M_{Nx}=1.01$ vs. $M_{base}=0.45$ & 0.002** & 0.839 \\
Reengagement rate & $M_{Nx}=0.37$ vs. $M_{base}=0.09$ & 0.008** & 0.688 \\
Avg. out degree & $M_{Nx}=0.98$ vs. $M_{base}=0.30$ & 0.004** & 0.781 \\
Max width & $M_{Nx}=66.93$ vs. $M_{base}=4.21$ & 0.001*** & 0.881 \\
LCC ratio & $M_{Nx}=0.29$ vs. $M_{base}=0.43$ & 0.021* & 0.638 \\
Num. components & $M_{Nx}=6.64$ vs. $M_{base}=4.14$ & 0.030* & 0.607 \\
Avg. betweenness & $M_{Nx}=0.00$ vs. $M_{base}=0.00$ & 1.000 & 0.141 \\ \bottomrule
\addlinespace[1ex]
\multicolumn{4}{l}{\small * $p < .05$, ** $p < .01$, *** $p < .001$} \\
\end{tabular}
\end{table}

\begin{figure}[h]
    \centering
    \includegraphics[width=1.05\linewidth]{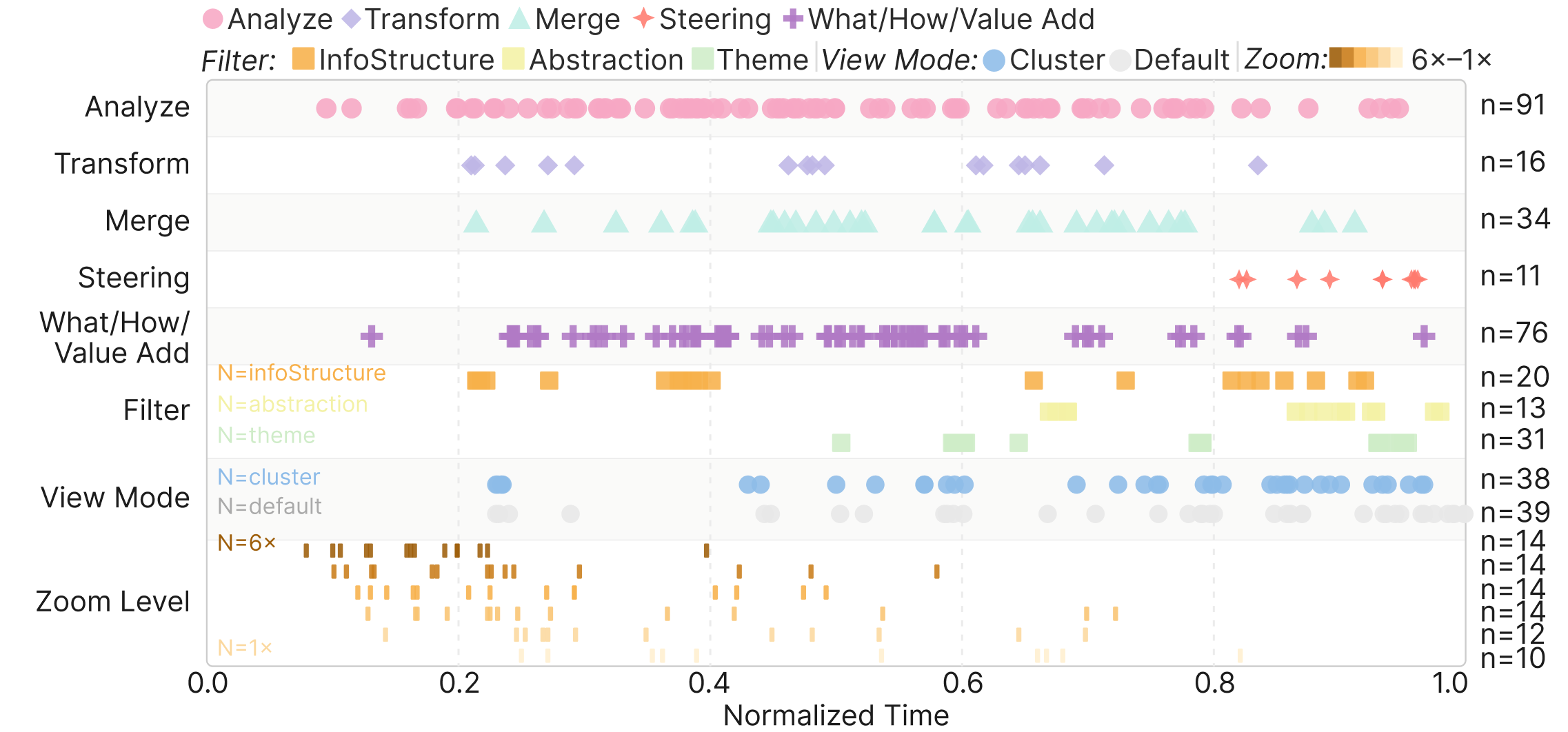}
    \caption{Distribution of interaction events across normalized time for 14 participants in NexusAI. Three phases emerged: early Analyze and \WHV{} Add, middle Merge and View Mode switching, and late Cluster, theme Filter, and Steering. Zoom-level patterns suggest navigation stabilized in the first half of the session.}
    \label{fig:Nexus timeline}
\end{figure}

\begin{figure*}[h]
    \centering
    \includegraphics[width=0.7\linewidth]{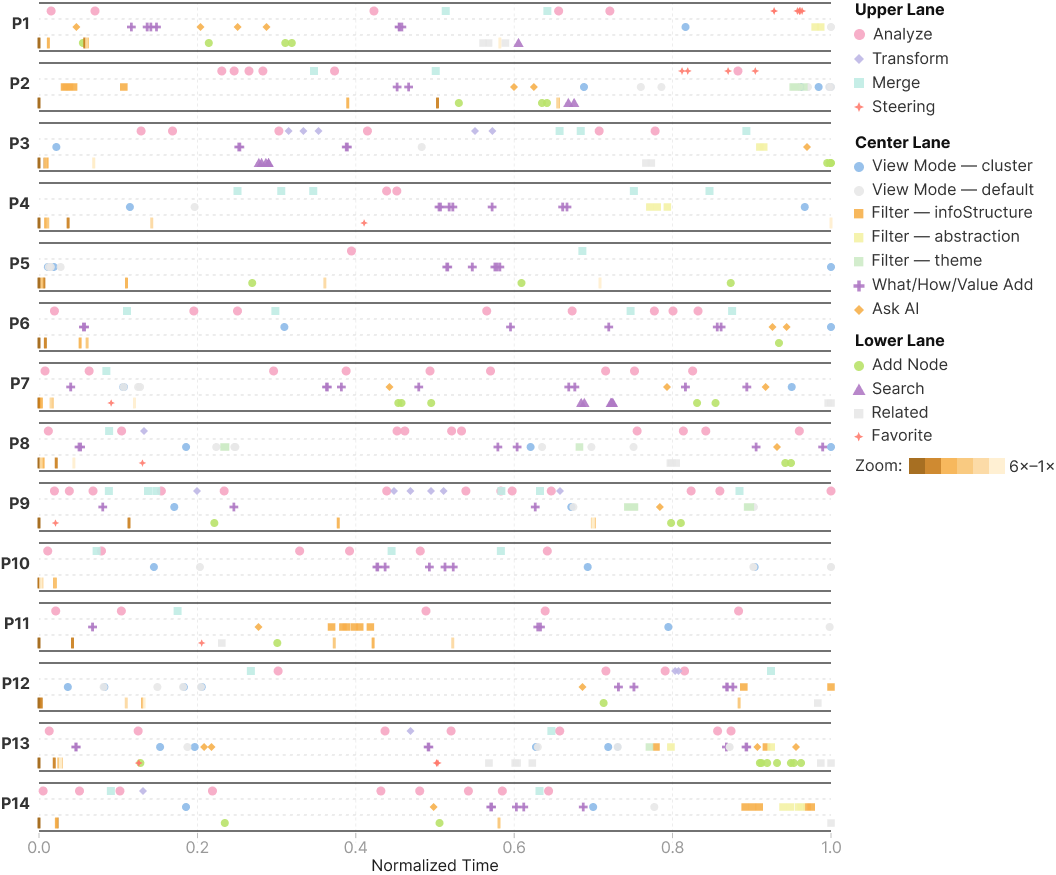}
    \caption{Interaction events analysis of 14 participants in a normalized time distribution.}
    \label{fig:log}
\end{figure*}

System logs are shown in Figure \ref{fig:Nexus timeline} and Figure \ref{fig:log}. 

\begin{acks}
To Robert, for the bagels and explaining CMYK and color spaces.
\end{acks}

\end{document}